\pdfoutput=1
\documentclass[aps,prb,twocolumn]{revtex4-1}

\usepackage{amsmath}
\usepackage{amssymb}
\usepackage{graphicx}
\usepackage{multirow}
\usepackage[usenames,dvipsnames]{xcolor}
\usepackage{siunitx}

\begin{document}

\title{Microscopic, first-principles model of strain-induced interaction in concentrated size-mismatched alloys}

\author{I.\ A.\ Zhuravlev}
\affiliation{Department of Physics and Astronomy and Nebraska Center for Materials and Nanoscience, University of Nebraska--Lincoln, Lincoln, Nebraska 68588, USA}

\author{J.\ M.\ An}
\affiliation{Department of Physics and Astronomy and Nebraska Center for Materials and Nanoscience, University of Nebraska--Lincoln, Lincoln, Nebraska 68588, USA}

\author{K.\ D.\ Belashchenko}
\affiliation{Department of Physics and Astronomy and Nebraska Center for Materials and Nanoscience, University of Nebraska--Lincoln, Lincoln, Nebraska 68588, USA}

\date{September 10, 2014}

\begin{abstract}
The harmonic Kanzaki-Krivoglaz-Khachaturyan model of strain-induced interaction is generalized to concentrated size-mismatched alloys and adapted to first-principles calculations. The configuration dependence of both Kanzaki forces and force constants is represented by real-space cluster expansions that can be constructed based on the calculated forces. The model is implemented for the fcc lattice and applied to Cu$_{1-x}$Au$_x$ and Fe$_{1-x}$Pt$_x$ alloys for concentrations $x=0.25$, 0.5, and 0.75. The asymmetry between the $3d$ and $5d$ elements leads to large quadratic terms in the occupation-number expansion of the Kanzaki forces and thereby to strongly non-pairwise long-range interaction. The main advantage of the full configuration-dependent lattice deformation model is its ability to capture this singular many-body interaction. The roles of ordering striction and anharmonicity in Cu-Au and Fe-Pt alloys are assessed. Although the harmonic force constants defined with respect to the unrelaxed lattice are unsuitable for the calculation of the vibrational entropies, the phonon spectra for ordered and disordered alloys are found to be in good agreement with experimental data. The model is further adapted to concentration wave analysis and Monte Carlo simulations by means of an auxiliary multi-parametric real-space cluster expansion, which is used to find the ordering temperatures. Good agreement with experiment is found for all systems except CuAu$_3$ (due to the known failure of the generalized gradient approximation) and FePt$_3$, where the discrepancy is likely due to the neglect of magnetic disorder.
\end{abstract}

\maketitle

\section{Introduction}

The standard theoretical framework for predicting phase diagrams and other thermodynamic properties of alloys requires an adequate representation of the formation enthalpy in the form of the Ising model with effective cluster interaction parameters. \cite{deFontaine} The construction of such models is a major goal of the first-principles alloy theory. \cite{deFontaine,Sanchez,Duc,Zunger,Walle,Ruban-review}

An important part of the effective configurational interaction in size-mismatched alloys comes from atomic relaxations. \cite{deFontaine,Krivoglaz,Khachaturyan} This strain-induced interaction is unavoidably long-range, because a local static disturbance in the atomic positions is propagated through the crystal by a response function with an anisotropic spectrum that is gapless thanks to the translational invariance. An adequate real-space cluster expansion of the strain-induced interaction is therefore often difficult to construct based on the structure inversion method.\cite{CW}

Two approaches have been widely used to represent the strain-induced interaction by an effective \emph{pairwise} configurational interaction: the mixed-basis cluster expansion with the so-called constituent strain (CS) contribution, \cite{Laks,Ozolins2,Blum} and the Kanzaki-Krivoglaz-Khachaturyan model (KKKM) assuming configuration-independent force constants and Kanzaki forces that are linear in occupation numbers.\cite{Matsubara,Kanzaki,Krivoglaz,Khachaturyan,deFontaine} The long-range character of the strain-induced interaction is reflected in the orientation-dependent discontinuity at the $\Gamma$ point in reciprocal space.
The CS approach determines the anisotropy of this singularity using the relaxation energies of coherent superlattices with infinitely separated parallel phase boundaries. The assumption of a purely pairwise strain-induced interaction in the entire concentration range appears to be an unavoidable and uncontrollable approximation in this approach. In the KKKM approach the full relaxation energy (and not just its long-range limit) is described by an effective pairwise interaction, but its assumptions restrict its quantitative applicability to the case of a dilute alloy with configuration-independent force constants. In the dilute limit, first-principles calculations can be performed using large supercells with isolated impurities. \cite{Ruban-2010,Ruban-2014} However, interactions obtained for a dilute alloy can not generally be extended to the concentrated case. Force constants in alloys were extensively studied in connection with phonon spectra and vibrational entropies, \cite{vdW-RMP,Fultz} and they often depend strongly on the configuration. As we will show below, the Kanzaki forces are strongly non-linear in occupation numbers in $3d$-$5d$ alloys, which we believe to be a generic situation.

The general structure of the effective strain-induced interaction in the harmonic approximation involves an inverse of the force constant matrix contracted on both sides with the vector representing the Kanzaki forces. This structure does not, in fact, depend on the simplifying assumptions made within the KKKM. However, if these assumptions are not satisfied, the effective interaction is by no means pairwise. \cite{Shirley,BPSV} A many-body representation of the Kanzaki forces based on first-principles calculations was reported for the Cu$_{0.75}$Au$_{0.25}$ alloy system.\cite{Shchyglo} However, the force constants were still assumed to be independent of the configuration, while the many-body representation of the Kanzaki forces, as we show below, did not include the most important term beyond the traditional KKKM.

An alternative effective tetrahedron method \cite{Ruban-tetra} assumes a strictly local but non-pairwise form for the interaction adapted to a particular crystal lattice. While useful in specific cases, particularly as an addition to the coherent potential calculations in which accurate forces are unavailable, this approximation is uncontrollable and does not aim to describe the singular part of the strain-induced interaction.

An important general feature of the problem should also be emphasized. The assumption of the continuity of the crystal lattice implies that all the ensuing predictions will correspond to coherent phase transformations. It is a general property of such transformations in phase-separating alloys that the conventional concept of an equilibrium phase diagram, with unique boundaries of single-phase regions, is inapplicable. \cite{Williams1,Williams2,CL,Larche,Johnson}
In particular, the concentrations of two phases in equilibrium depend not only on temperature, but also on the overall composition of the alloy. This dependence appears because the state of strain of the phases in equilibrium depends on their molar fractions; the common tangent construction is invalid for coherently stressed solids because the elastic free energy is not a sum of contributions from the two phases. The convex hull in the formation enthalpy diagram likewise loses its clear meaning. It therefore appears to be unavoidable that a physically meaningful configurational Hamiltonian of a coherently strained alloy should depend on its overall composition treated as a macroscopic parameter. The long-range character of the strain-induced interaction presents an obstruction to any local representation of this dependence. The commonly used methods mentioned above do not respect this general feature. In particular, although the pairwise interaction potential in the CS approach does depend on the concentration, in practical applications it is usually combined with a semi-grand canonical Monte Carlo simulation, which relaxes the condition of fixed overall concentration and yields a universal phase diagram.

In this paper we generalize the KKKM based on controlled cluster expansions for both Kanzaki forces and force constants that can be constructed using first-principles calculations. This configuration-dependent lattice deformation model (CLDM) is designed to capture the \emph{many-body} strain-induced interaction on all length scales. The configurational Hamiltonian by construction corresponds to a fixed overall composition.

The paper is organized as follows. The CLDM model is formulated in Section \ref{sec:model}, and Section \ref{methods} explains the computational methods. The subsequent sections report on the application of CLDM to Cu-Au and Fe-Pt alloys. The cluster expansions for the Kanzaki forces and force constants are described in Sections \ref{sec:forces} and \ref{sec:fc}, respectively. The relaxation energies predicted by CLDM are discussed in Section \ref{Harmrelaxationen}. Section \ref{striction} deals with the role of striction (homogeneous strain), and Section \ref{anharmonic} with the anharmonicity. The phonon spectra for ordered and disordered alloys are calculated and compared with experiment in Section \ref{sec:phonons}. The auxiliary cluster expansion for the relaxation energy is presented in Section \ref{auxce}, and the the effective pair interaction for a nearly random alloy is analyzed in Section \ref{sec:veff}. The phase transitions are found using Monte Carlo simulations in Section \ref{mcresults}. Finally, the conclusions are drawn in Section \ref{conclusions}.

\section{Configuration-dependent lattice deformation model}\label{sec:model}

In this section we describe the configuration-dependent lattice deformation model (CLDM), which generalizes the Kanzaki-Krivoglaz-Khachaturyan model to concentrated alloys and may be readily constructed using first-principles data. The accuracy of the model can be systematically improved. The present formulation is restricted to the harmonic approximation; the errors due to anharmonicity are discussed later in Section \ref{anharmonic}.

\subsection{Harmonic Hamiltonian of a concentrated alloy}

Following the standard approach, \cite{deFontaine} we start from the separation of the formation enthalpy of an ordered structure in two contributions: the ``chemical'' part $H_{chem}$ corresponding to all atoms fixed at ideal positions of the parent lattice, and the relaxation part $H_{rel}$ associated with displacements away from these ideal positions:
\begin{equation}\label{Htot}
    H\left(\Sigma,\{\mathbf{u}\}\right)=H_{chem}(\Sigma)+H_{rel}\left(\Sigma,\{\mathbf{u}\}\right),
\end{equation}
where we used $\Sigma$ to denote the configuration of the (generally multi-component) alloy, i.\ e.\ occupations of all lattice sites by atoms of different types, and $\{\mathbf{u}\}$ to denote the displacements of all atoms from their ideal positions.

While $H_{chem}$ usually presents no problems for a conventional cluster expansion, our focus is on the difficult part $H_{rel}$.
In the harmonic approximation it can be written as:
\begin{equation}\label{Hrel}
    H_{rel}=-\sum\limits_{i}\mathbf{u}_i \mathbf{F}_{i}(\Sigma)+\frac12\sum_{ij}\mathbf{u}_i
    \hat A_{ij}(\Sigma)\mathbf{u}_j
\end{equation}
where $\mathbf{F}_i$ is the force acting on the atom at site $i$ in the unrelaxed state, and $\hat A_{ij}$ is the configuration-dependent force constant matrix.
Using the fact that the energy must be invariant with respect to global translations after any deformation, one can show
\begin{equation}\label{Ainv}
    \sum_j \hat A_{ij}(\Sigma)=0 \quad  \textrm{for any $\Sigma$.}
\end{equation}

To proceed, we will construct cluster expansions for the forces $\mathbf{F}_{i}(\Sigma)$ and force constants $A_{ij}(\Sigma)$. It is natural to expect that these expansions should quickly converge in real space. A first-principles cluster expansion of the Kanzaki forces for Cu$_3$Au alloys was undertaken by Shchyglo \emph{et al.} \cite{Shchyglo} based on the fit of the total energies, but, as we will see below, one of the dominant terms was missed, and the resulting expansion does not provide an adequate representation of the forces. Configuration-dependent force constants have been extensively studied from first principles to understand the role of vibrational entropy in the thermodynamics of phase transitions.\cite{vdW-RMP} It was found that the force constants in the equilibrium configuration depend strongly on the bond lengths due to anharmonicity, and this dependence is often taken into account by introducing explicit distance dependence. In the present treatment the situation is simpler, because the relaxation energy (\ref{Hrel}) is defined with respect to the ideal lattice, in which all the bond lengths are equal. Thus, based on the earlier studies one may expect that a small number of configuration-dependent terms is sufficient to adequately represent the force constants in (\ref{Hrel}), and below we will show this to be the case in Cu-Au and Fe-Pt alloys. We will also see that the anharmonic terms have a relatively small effect on the relaxation energies (particularly for the configurations where these energies are not too large), even though the harmonic force constants referenced from the equilibrium configuration may deviate significantly from those in (\ref{Hrel}).

\subsection{Cluster expansion of the Kanzaki forces}

The force term can be written as
\begin{equation}\label{Fi}
    \mathbf{F}_i(\Sigma)=\sum\limits_{a,P}    n_{ia} n_P \mathbf{F}_{ia,P}
\end{equation}
where $\mathbf{F}_{ia,P}$ is the contribution to the force acting on site $i$
occupied by atom of type $a$ due to the occupation of a cluster $P$, represented
by the projection operator $n_P=\prod_{jb\in P} n_{jb}$. Here $n_{jb}$ is the
occupation number of site $j$ by component $b$, and the definition of cluster
$P$ includes the set of sites and atom types occupying them. ($n_{ia}=1$ if site $i$ is occupied by atom type $a$ or 0 otherwise.) It is
assumed that $P$ does not contain $i$ in (\ref{Fi}).

In an $n$-component alloy there are $n-1$ independent occupation numbers for each site. One of the original variables $n_{ia}$ may be eliminated using $\sum_a n_{ia}=1$, or one can introduce a different set of $n-1$ variables linearly related to $n_{ia}$. For a binary alloy A-B it is convenient to use Ising variables $\sigma_i=n_{iA}-n_{iB}$; in general there are $\sigma_{ic}$ of $n-1$ flavors $c$. Expressing all the occupation numbers in (\ref{Fi}) through $\sigma_{ic}$ (and omitting the summation over flavors needed in the multi-component case), we obtain
\begin{equation}\label{Fi2}
    \mathbf{F}_i=\sum\limits_{P}\mathbf{F}_{iP}\sigma_P
\end{equation}
where the cluster $P$ now can contain site $i$, and $\sigma_P=\prod_{j\in P} \sigma_{j}$.
The sum in (\ref{Fi2}) does not include the empty cluster, because this term would correspond to a macroscopic force acting on the entire crystal.

Since for any set of $\sigma_P$ the total energy should be translationally invariant, the forces satisfy the condition
\begin{equation}\label{transl}
    \sum_{i} \mathbf{F}_{iP}=0
\end{equation}
To each cluster $P$ one can formally assign a displacement $\mathbf{u}_P$. For a single-atom cluster this should obviously be the displacement of the corresponding atom. For multi-atom clusters one can, for example,
take the average displacement of the atoms included in $P$. Using (\ref{transl}), we can then rewrite the linear term in
(\ref{Hrel}) as
\begin{equation}\label{Hrel2}
    H_K=-\sum\limits_{iP}(\mathbf{u}_i-\mathbf{u}_P)\mathbf{F}_{iP}\sigma_P
\end{equation}
or
\begin{equation}\label{Hrel3}
    H_K=-\frac12\sum\limits_{ij}(\mathbf{u}_i-\mathbf{u}_j)\mathbf{F}_{ij}(\Sigma)
\end{equation}
where
\begin{equation}\label{FK}
    \mathbf{F}_{ij}(\Sigma)=2\sum_{P\ni j}\mathbf{F}_{iP}\frac{\sigma_P}{N_P}
\end{equation}
with a sum over all clusters $P$ containing site $j$, and $N_P$ is the number of sites in $P$.

In view of the structure of $(\ref{Hrel3})$, the forces may always be taken to be antisymmetric with respect to indices $i$, $j$, i.\ e.\ $\mathbf{F}_{ij}(\Sigma)=-\mathbf{F}_{ji}(\Sigma)$. This property will be assumed in the remainder of this section; it means that the quantity $\mathbf{F}_{ij}(\Sigma)$ can be interpreted as the force exerted by site $j$ on site $i$ in the configuration $\Sigma$, which obeys Newton's third law. However, when we consider specific alloy systems in the following, it is more convenient to introduce the expansion directly for $\mathbf{F}_i$, simply requiring that the condition (\ref{transl}) is satisfied.

In addition to the translational invariance (\ref{transl}), the condition of rotational invariance must be imposed, which requires that the total torque acting on the crystal vanishes in any configuration:
\begin{equation}
\sum_i\mathbf{F}_{iP}\times \mathbf{r}_i=0.
\label{rotation}
\end{equation}

Let us further discuss the symmetry properties of the expansion. Consider a particular term $\mathbf{F}_{iP}$ in the expansion for $\mathbf{F}_i$. The vector $\mathbf{F}_{iP}$ can always be chosen to be invariant under the subgroup $G_{iP}$ of the point group $G_i$ of site $i$ that leaves the cluster $P$ invariant (while possibly permuting some of its sites). Therefore, if we consider the natural representation of $G_{iP}$ in $R^3$, the dimension of its invariant subspace $V_{iP}$ gives the number of components of $\mathbf{F}_{iP}$ that may be independently varied. Introducing a basis $\mathbf{e}^\nu_{iP}$ that spans $V_{iP}$, we write
\begin{equation}
\mathbf{F}_i=\sum_{P,\nu} f^\nu_{iP} \sigma_P \mathbf{e}^\nu_{iP},
\label{fi2}
\end{equation}
and coefficients $f^\nu_{iP}$ play the role of the fitting parameters. (Note that in addition to crystal symmetry they are subject to conditions (\ref{transl}) and (\ref{rotation}).) Consider now two clusters $P$ and $\bar P$ such that $i\notin P$ and $\bar P = P\cup\{i\}$. Since they differ only in the presence of site $i$, the groups $G_{iP}$ and $G_{i\bar P}$ are identical, and so are the corresponding invariant subspaces $V_{iP}$ and $V_{i\bar P}$. We are therefore free to select the same basis $\mathbf{e}^\nu_{iP}$ for $V_{iP}$ and $V_{i\bar P}$ and rewrite (\ref{fi2}) as
\begin{equation}
\mathbf{F}_i={\sum_{P,\nu}}^\prime \left(f^\nu_{iP}+\sigma_i \bar f^\nu_{iP}\right) \sigma_P \mathbf{e}^\nu_{iP},
\label{fi3}
\end{equation}
where the prime indicates that the summation is taken over clusters $P$ that do not contain $i$. This representation is very convenient in applications. Note that the terms $\bar f^\nu_{iP}$ were not included in Ref.\ \onlinecite{Shchyglo}. We will see below that the nearest-neighbor quadratic term $\bar f_1$ is comparable in magnitude to the conventional Kanzaki term $f_1$, representing a large asymmetry of the forces with respect to the interchange of the two components ($\sigma_i\to-\sigma_i$).

\subsection{Total energy of an ordered supercell}

We will be dealing with relaxed or unrelaxed configurations of ions in supercells, for which one can calculate energies, forces, and strains from first principles. For each supercell of any particular ordering the
displacement of basis atom $i$ can be written as
\begin{equation}\label{displ}
    \mathbf{u}_i=u_{\alpha\beta}(R^\beta_i-R^\beta_0)+\mathbf{w}_i
\end{equation}
where $u_{\alpha\beta}$ is the symmetric strain tensor, $\mathbf{R}_0$ is the coordinate of the origin, and $\mathbf{w}_i$ is the periodic (internal) part
of the displacement of basis atom $i$ (determined with respect to the homogeneously deformed ideal lattice). The general deformation can be written
as (\ref{displ}), because the antisymmetric part of the homogeneous deformation is an inconsequential uniform rotation of the lattice.

Substituting the displacement (\ref{displ}) into (\ref{Hrel}) and taking into
account (\ref{Hrel3}), we find the general expression
\begin{align}\label{Hrel5}
    H_{rel}&=\frac12\sum\limits_{ij}\mathbf{w}_{ij}\mathbf{F}_{ij}(\Sigma)
    +\frac12\sum_{ij}\mathbf{w}_{i}\hat A_{ij}(\Sigma)\mathbf{w}_{j}\nonumber\\
    &+\frac12u_{\alpha\beta}\sum\limits_{ij}\mathbf{F}^\alpha_{ij}(\Sigma)R^\beta_{ij}
    +u_{\alpha\beta}\sum\limits_{ij}w^\gamma_{i}A^{\gamma\alpha}_{ij}(\Sigma)R^\beta_{ij}\nonumber\\
    &+\frac12 u_{\alpha\gamma}
    u_{\beta\delta}\sum_{ij}A^{\alpha\beta}_{ij}(\Sigma)R^\gamma_{i}R^\delta_{ij},
\end{align}
where $\mathbf{R}_{ij}=\mathbf{R}_j-\mathbf{R}_i$.
If the total energy corresponding to the volume of one supercell is required, the sum over $i$ in (\ref{Hrel5}) should be taken over all the basis sites in the supercell. (Notice that in the last term $R^\gamma_i$ then becomes the internal basis vector.) The sum over $j$ is, however, taken over all the sites in the infinite lattice, whose range is limited by the range of the Kanzaki forces and the dynamical matrix. Energy per site is obtained by
dividing by the number of basis sites.

The configurational dependence of the force constants leads to the presence of a bilinear coupling between the homogeneous strain $u_{\alpha\beta}$ and lattice relaxations $\mathbf{w}_i$. It implies that the forces change under a homogeneous strain deformation even if the atoms remain at their ideal lattice positions. Conversely, the stress tensor changes as the atomic positions are relaxed. These features are obvious in first-principles calculations but absent from from traditional KKKM based on the assumption of configuration-independent force constants.

\subsection{Equilibrium conditions and strain-induced interaction}

The equilibrium conditions are obtained by minimizing $H_{rel}$ with respect to $\mathbf{w}_i$ and $u_{\alpha\beta}$. This means that the forces:
\begin{align}\label{forces}
    F^\alpha_i=\sum_j\left[F^\alpha_{ij}(\Sigma)-
    A^{\alpha\beta}_{ij}w^\beta_{ij}-u_{\gamma\delta}A^{\alpha\gamma}_{ij}(\Sigma)R^\delta_{ij}\right]
\end{align}
where $\mathbf{w}_{ij}=\mathbf{w}_j-\mathbf{w}_i$, and the stress tensor
\begin{align}\label{stress}
    V\sigma_{\alpha\beta}
    &=\frac14\sum\limits_{ij}F^\alpha_{ij}(\Sigma)R^\beta_{ij}
    +\frac12\sum\limits_{ij}w^\gamma_{i}A^{\gamma\alpha}_{ij}(\Sigma)R^\beta_{ij}\nonumber\\
    &+\frac12u_{\nu\delta}\sum_{ij}A^{\alpha\nu}_{ij}R^\beta_{i}R^\delta_{ij}+\{\alpha\leftrightarrow\beta\}=0
\end{align}
vanish in equilibrium. The term with $R_i^\beta$ in $\sigma_{\alpha\beta}$ is independent on the choice of the origin thanks to the invariance properties of the dynamical matrix. Note that the quantity $\sigma_{\alpha\beta}$ is equal to the physical stress tensor only if the internal displacements $\mathbf{w}_i$ correspond to vanishing forces.

In the harmonic approximation the equilibrium conditions derived above are linear in $\mathbf{w}_i$ and $u_{\alpha\beta}$. For any ordered configuration of the alloy (i.\ e.\ with any finite unit cell) the number of variables and equations is finite. The solution is unique up to a homogeneous translation, which can be explicitly excluded. By solving the linear equilibrium conditions and substituting the solutions back into $H_{rel}$, one can calculate the relaxation energy of the given configuration as a function of the parameters appearing in the cluster expansions for the Kanzaki forces and force constants.

In this work we concentrate on evaluating the relaxation energy for alloys with concentration near a certain chosen value. Further, in calculations based on CLDM we exclude the contribution of the homogeneous strain induced by ordering (i.\ e.\ striction) by setting $u_{\alpha\beta}=0$. (The role of striction for the alloys considered here is discussed in Section \ref{striction}.) The equilibrium atomic displacements under this restriction are
\begin{equation}
W=\hat A^{-1} F
\label{WAF}
\end{equation}

For a supercell with $N$ atoms, $F$ and $W$ are $3N$-dimensional column vectors representing all the force and displacement components. Although the matrix $\hat A$ has three zero eigenvalues corresponding to homogeneous translations, its inversion is easily conditioned by adding a fictitious finite stiffness associated with the displacement of the whole crystal. As a result, the displacements of all basis sites obtained from (\ref{WAF}) automatically add up to zero. Substituting $W$ in $H_{rel}$, one finds the relaxation energy
\begin{equation}
E_{rel}(\Sigma)=-\frac12\sum_{ij}\mathbf{F}_i(\Sigma)\hat A^{-1}_{ij}(\Sigma)\mathbf{F}_j(\Sigma),
\label{Erel}
\end{equation}
where the matrix $\hat A$ should be conditioned before inversion as explained above. Eq.\ (\ref{Erel}) is \emph{exact} in the harmonic approximation.

Since both $\mathbf{F}$ and $\hat A$ are represented by cluster expansions, the resulting interaction is by no means pairwise, contrary to the commonly used form of KKKM. In fact, if the configuration dependence of the force constants is taken into account, it contains interactions or arbitrarily high order. One can, in principle, expand $\hat A^{-1}$ in the occupation variables with respect to some ``average-alloy'' reference point and obtain an infinite sequence of many-body interactions, each of which is long-range and inherits the long-distance singularity of the dynamical matrix. On the other hand, even in the dilute alloy limit the configurational dependence of the force constants is important. \cite{Krivoglaz} This can be immediately seen from the extreme case when the force constants binding the impurity atom to its neighbors are much larger compared to the host material. In this limit the equilibrium displacements of the neighbors induced by the Kanzaki forces, and thereby the strain-induced interaction, are suppressed.

As we will see in the following, in $3d$-$5d$ alloys like Cu-Au and Fe-Pt the asymmetry between the components leads to large two-body terms in the Kanzaki forces and thereby to significant non-pairwise strain-induced interactions --- even if the configurational dependence of $\hat A$ is disregarded. This feature reveals the intrinsic limitation of the methods designed to describe the strain-induced interaction using pairwise configurational interaction.

\section{Computational methods and input datasets}\label{methods}

For applications we have chosen Cu--Au and Fe--Pt as typical and well-studied $3d$-$5d$ alloy systems. Due to a fairly large size mismatch (13\% and 10\%, respectively), strain-induced interaction in these systems is comparable with the chemical contribution and competes with it, which makes them appropriate for the application of CLDM.

In this work we focus on constructing an accurate representation of the lattice relaxation energy at a fixed overall composition of the alloy. As mentioned in the Introduction, this representation is generally suitable for the description of coherent phase transformations in combination with thermodynamic simulations in the canonical ensemble. For ordering phase transitions occurring with a narrow two-phase region (e.\ g.\ close to a point of equal concentration) or through a second-order transition the requirement of a canonical ensemble simulation is not critical. The Cu-Au phase diagram has points of equal concentration near CuAu and Cu$_3$Au compositions, while Fe-Pt alloys have points of equal concentration near Fe$_3$Pt, FePt, and FePt$_3$ compositions. The low-temperature phases have L1$_0$ and L1$_2$ orderings. The Cu-Au system also has an L1$_2$-ordered CuAu$_3$ phase terminating at a peritectic point. Thus, for both alloy systems construct effective configurational Hamiltonians at three concentrations of 25\%, 50\%, and 75\%.

The input sets for 50\% concentration consisted of all 27 ordered structures with up to 6 atoms in the unit cell. For the $3:1$ compositions the input sets included all 7 structures with 4-atom unit cells and 25 structures with 8-atom unit cells. The latter included the total of 9 superlattices with A$_6$B$_2$, A$_5$BAB, and A$_4$BA$_2$B stackings along [100], [110], and [111] directions, which were used in Ref.\ \onlinecite{Shchyglo}. In some calculations we also added a 16-atom special quasirandom structure (SQS16) in each set.

Total energy calculations were performed using the projected augmented wave (PAW) method \cite{Bloechl,VASP-PAW} and the PBEsol \cite{PBEsol} exchange-correlation functional as implemented in the VASP  package. \cite{VASP} The choice of PBEsol was motivated by the fact that it gives the lattice parameter of Au in much better agreement with experiment compared to PBE. An energy cutoff of 350 eV was used for the plane-wave basis set, and the reciprocal-space integration was performed on a $\Gamma$-centered mesh with a density equivalent to at least $16\times16\times16$ points in the cubic Brillouin zone of the parent fcc lattice. Structural optimizations were performed using the conjugate-gradient algorithm and the Methfessel-Paxton smearing scheme.

The CLDM parameters are determined from the calculated forces for all input structures calculated either with ideal atomic positions or with small displacements of individual atoms, as explained in the following two sections. All these calculations were performed with an ideal shape and volume of the unit cell for the following lattice parameters: 3.688 \AA\ for 25\% Au, 3.819 and 3.897 \AA\ for 50\% Au, and 3.955 \AA\ for 75\% Au in the Cu-Au system, and 3.679 \AA, 3.769 \AA, and  3.850 \AA\ for the same concentrations of Pt in the Fe-Pt system. Note that we used two different lattice parameters for the Cu$_{0.5}$Au$_{0.5}$ system in order to check the sensitivity of CLDM to this choice. The first value 3.819 \AA\ was chosen somewhat arbitrarily (as they were for the other systems), while the other value 3.897 \AA\ was designed to minimize the volume relaxation energies. This was done as follows. For each input structure $\nu$ the energy per atom $E_\nu$ was calculated and fitted as a function of $a$. Then the sum $\sum_\nu (E_\nu-E^0_\nu)$ was minimized with respect to $a$. We found that the choice of the lattice parameter has a small effect on the model parameters and thermodynamic properties, showing that a reasonably crude choice within 1-2\% of equilibrium is acceptable.

\section{Kanzaki forces}\label{sec:forces}

In this and subsequent sections we use first-principles data to construct CLDM for Cu-Au and Fe-Pt alloys at three concentrations: 0.25, 0.5, and 0.75. Our first step is to parameterize the Kanzaki forces based on the general expression (\ref{fi3}). For each dataset we fitted the parameters of (\ref{fi3}) to the set of all Hellman-Feynman forces calculated for the unrelaxed ordered structures in the dataset. (This information is already available in a conventional cluster-expansion procedure, and we have checked that the calculated forces were sufficiently converged with respect to the energy cutoff and $k$-point density.) Our strategy was to start with a list of 16 symmetry-respecting terms in the expansion, and for each dataset with a given alloy composition to include only those terms that appreciably reduced the cross-validation score for the forces. This procedure resulted in 7-8 terms for each dataset.

The simplest cluster type is a single site $P=\{j\}$. If $j$ is a nearest neighbor of $i$ in the fcc (or bcc) lattice, the invariant subspace $V_{ij}$ is one-dimensional, and the single basis vector $\mathbf{e}_{ij}$ is parallel to $\mathbf{r}_{ij}$ (i.\ e.\ central force). The same applies to second, fourth, and fifth-nearest neighbors in the fcc lattice, but for third-nearest neighbors the invariant subspace is two-dimensional. We introduced an orthogonal basis for this two-dimensional subspace, with one of the basis vectors $\mathbf{e}_{ij}$ parallel to $\mathbf{r}_{ij}$ and representing a central force component, and the other basis vector $\mathbf{e}^\perp_{ji}$.
Thus, 10 of the 16 trial terms represent central forces between pairs of atoms up to fifth-nearest neighbors, and one more term describes the non-central force from the third-nearest neighbor. The contribution of all these terms to $\mathbf{F}_i$ is
\begin{equation}
\mathbf{F}^{(1)}_i=\sum_n \left(f_n+\bar f_n\sigma_i\right)\sum_{j\in S_n}\sigma_j\,\mathbf{e}_{ji}+f'_3\sum_{j\in S_3}\sigma_j\,\mathbf{e}^\perp_{ji}
\end{equation}
where $\sigma_i=1$ for $5d$-elements and $-1$ for $3d$-elements, and $S_n$ is the set of sites in the $n$-th coordination sphere of $i$. Note that the non-central term with $\sigma_i\sigma_j$ would violate the condition (\ref{rotation}) and is therefore excluded.

\begin{equation}
\mathbf{F}_i=\sum_{j=nn(i)} \left(f_1+\bar f_1\sigma_i\right)\sigma_j\,\mathbf{e}_{ji}
\end{equation}

The remaining five terms correspond to coplanar forces from two-site clusters $P=\{j,k\}$:
\begin{align}
\mathbf{F}^{(2)}_i=\sum_{jk}[(f_t+\sigma_i\bar f_t)\eta_{ijk}+(f_l+\sigma_i\bar f_l)\zeta_{ijk}\nonumber\\
+f_r\xi_{ijk}]\sigma_j\sigma_k\mathbf{e}^i_{jk}
\end{align}
where $\mathbf{e}^i_{jk}$ is a unit vector pointing from the midpoint between $j$ and $k$ towards $i$, and $\eta_{ijk}$, $\zeta_{ijk}$ and $\xi_{ijk}$ are projectors selecting specific cluster shapes. Namely, $\eta_{ijk}=1$ only if $i$, $j$ and $k$ make a triangle of nearest neighbors; $\zeta_{ijk}=1$ only if $i$, $j$, $k$ form a two-link straight chain of nearest neighbors with $i$ at an end; and $\xi_{ijk}=1$ only if $j$ and $k$ are both nearest neighbors of $i$ and each other's second-nearest neighbors (otherwise these factors are 0). Note that a tentative term with $\bar f_r\xi_{ijk}\sigma_i\sigma_j\sigma_k$ would not be allowed without counterterms to enforce translational invariance.

\begin{figure*}[hbt]
\begin{center}
\includegraphics[width=0.45\textwidth]{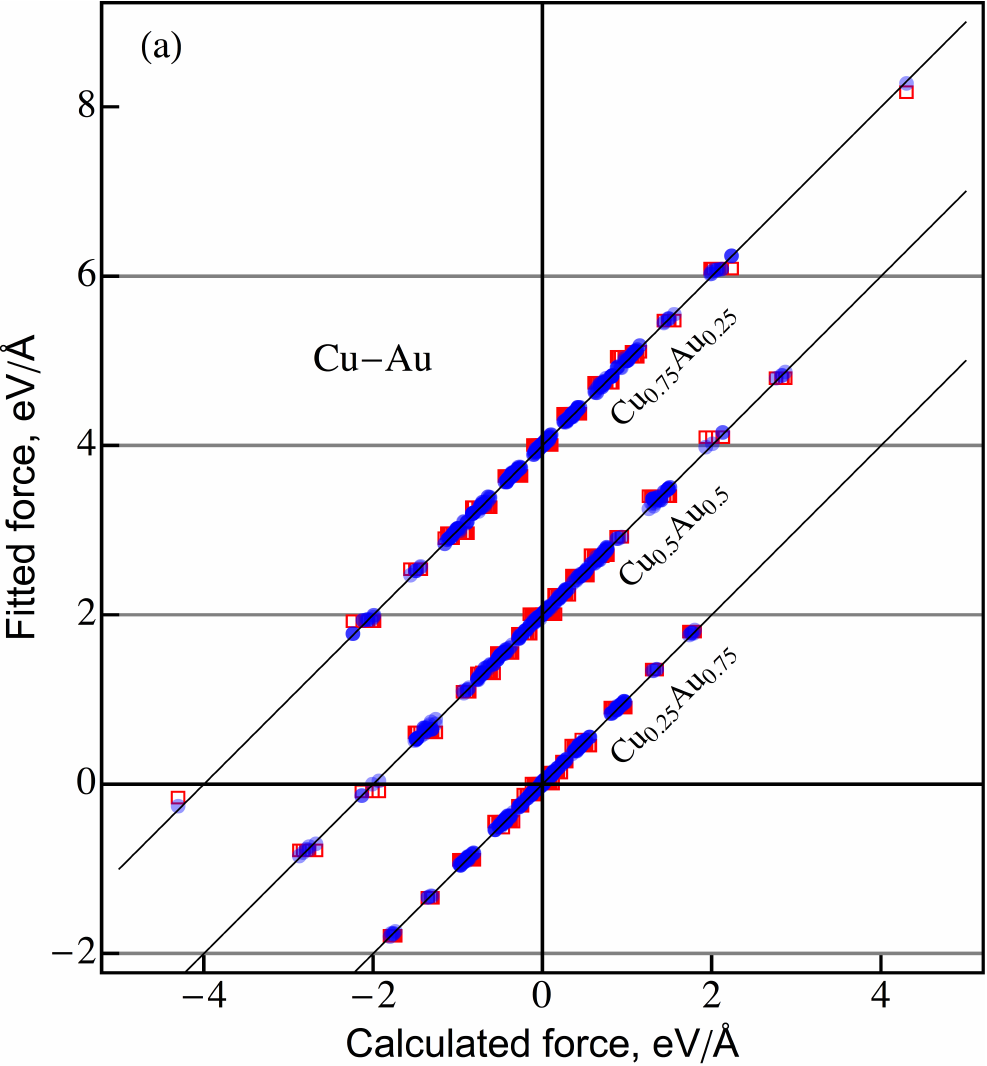}\hfil\includegraphics[width=0.45\textwidth]{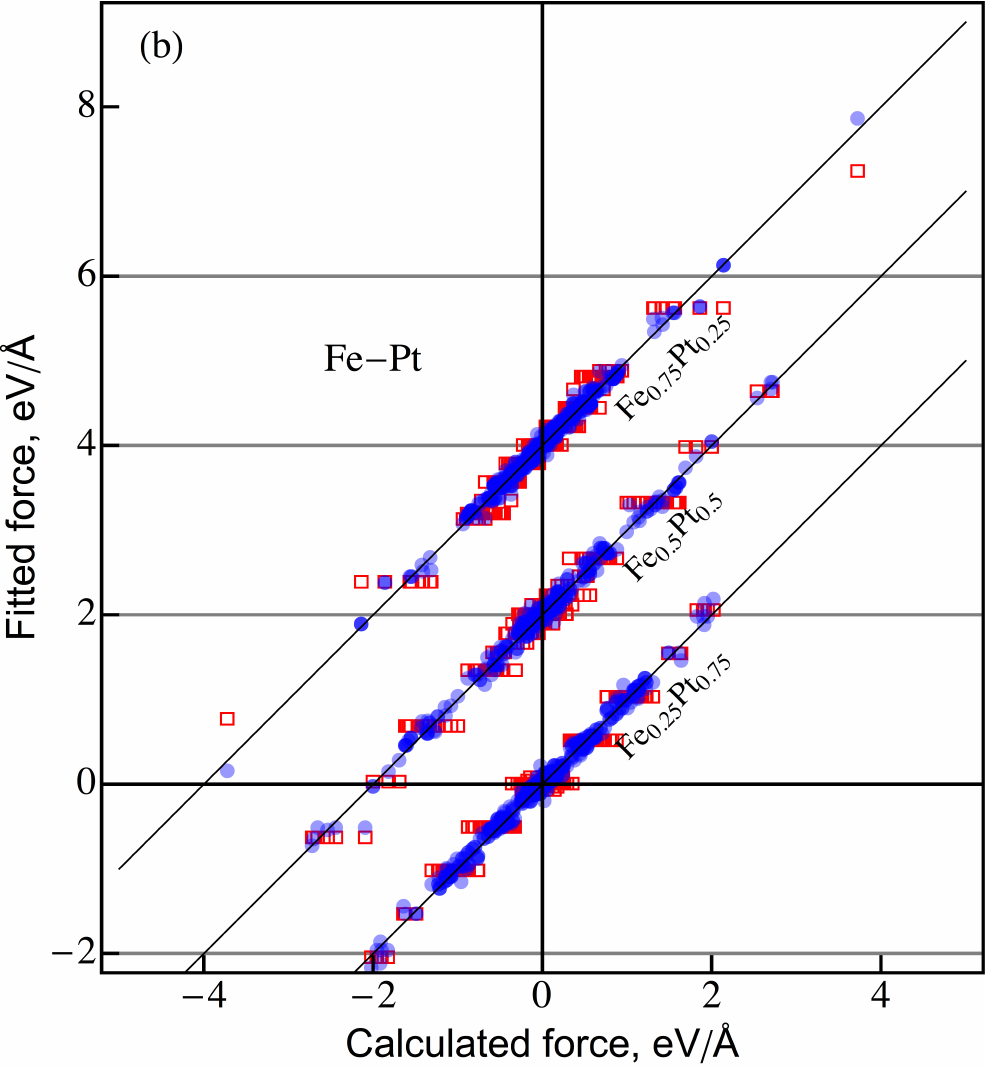}
\end{center}
\caption{(Color online) Fitting of the calculated forces to many-body Kanzaki force models. Each Cartesian component for each site is represented by one data point. Data sets for different concentrations are shifted by 2 units along the $y$ axis. Red open squares: Two-parameter model with nearest-neighbor Kanzaki forces. Blue opaque circles: complete model, see text and Table \ref{tab:forces}.}
\label{fig:forces}
\end{figure*}

The resulting sets of fitted parameters are listed in Table \ref{tab:forces} below the horizontal line, and the quality of the fits can be inferred from Fig.\ \ref{fig:forces} (filled circles). For all six systems the expansion is dominated by two nearest-neighbor terms $f_1$ and $\bar f_1$. For comparison, separate fits including only these two parameters are also included in Table \ref{tab:forces} (above the horizontal line) and in Fig.\ \ref{fig:forces} (empty squares).
It can be seen that these two-parameter fits are already reasonably good, particularly for the Cu-Au alloys, but the inclusion of additional parameters significantly improves the quality of the fit.
The set of fitted parameters for Cu$_{0.5}$Au$_{0.5}$ with the optimized lattice constant is similar to the one shown in Table \ref{tab:forces}.

\begin{table}
\caption{Parameters of the cluster expansion for the Kanzaki forces (units of eV/\AA). A horizontal line separates two different fittings.}
\begin{tabular}{|c|S|S|S|S|S|S|}
\hline
    & {CuAu$_3$} & {CuAu} & {Cu$_3$Au} & {FePt$_3$} & {FePt} & {Fe$_3$Pt} \\
\hline
 $f_1$ & 0.205 & 0.329 & 0.498 & 0.195 & 0.273 & 0.363 \\
 $\bar f_1$ & 0.113 & 0.167 & 0.238 & 0.168 & 0.193 & 0.209 \\
\hline
 $f_1$ & 0.200 & 0.323 & 0.502 & 0.206 & 0.262 & 0.373 \\
 $\bar f_1$ & 0.110 & 0.166 & 0.237 & 0.157 & 0.182 & 0.247 \\
 $f_2$ & 0.023 & 0.031 & 0.041 & 0.036 & 0.050 & 0.061 \\
 $\bar f_2$ & 0.003 & 0.005 & 0.008 &  &  &  \\
 $f_3$ & 0.003 & 0.005 & 0.006 & 0.015 & 0.013 &  \\
 $f_4$ & -0.011 & -0.011 & -0.010 & -0.035 & -0.020 &   \\
 $f_t$ & 0.003 & 0.004 & 0.008 &  &  & 0.015 \\
 $\bar f_t$ &  &  &  &  &  & 0.031 \\
 $f_l$ & -0.004 &  &  & -0.032 & -0.026 & -0.009 \\
 $f_r$ &  &  &  & 0.011 & 0.022 & 0.018 \\
\hline
\end{tabular}
\label{tab:forces}
\end{table}

An important feature obvious from Table \ref{tab:forces} is that the $\bar f_1$ term is comparable with $f_1$: for Cu-Au alloys $\bar f_1\approx 0.5f_1$, and for Fe-Pt $\bar f_1\approx 0.7 f_1$, and in both cases the $\bar f_1/f_1$ ratio decreases slightly with increasing concentration of the $3d$-element. The large value of this ratio indicates that the force acting on an atom placed at a particular site of a lattice with a certain ordering depends strongly on the identity of this atom. Specifically, in the model with only $f_1$ and $\bar f_1$ non-zero, the force acting on a $3d$ element at a particular site is proportional to $f_1-\bar f_1$, and if a $5d$-element takes its place then this factor changes to $-(f_1+\bar f_1)$. A similar relation applies to the forces acting on the atoms of $3d$ and $5d$-elements in a symmetric superlattice at 50\% concentration. For example, in the A$_2$B$_2$ [001] superlattice the calculated forces acting on Cu and Au atoms are 0.88 and 2.83 eV/\AA, respectively, and for the Fe-Pt superlattice they are 0.49 and 2.72 eV/\AA. This large asymmetry is properly described by our model. On the contrary, the model of Ref.\ \onlinecite{Shchyglo} for Cu$_3$Au does not include any of the $\bar f_{iP}$ parameters in Eq.\ (\ref{fi2}), and the force acting on site $i$ does not depend at all on its occupation. Therefore, that model (which was fitted to the relaxation energies of a few ordered structures and is dominated by the $f_1$ term) does not give an adequate representation of the Kanzaki forces.

\section{Force constants}\label{sec:fc}

Configuration-dependent force constants have attracted much attention, motivated mainly by the need to understand the effects of vibrational entropy on the thermodynamics of phase transitions, as well as by the general interest in the effects of ordering on the vibrational spectra.\cite{vdW-RMP,Fultz} However, to our knowledge, they have not been used in the studies of the strain-induced interaction. Usually a configuration-independent force constant matrix is introduced and fitted to provide some reasonable elastic moduli. Here we go beyond this approximation and construct a configuration-dependent representation of the force constant matrix using first-principles input data. Note that the definition of the force constants suitable for the evaluation of the relaxation energies is different from those designed to describe the vibrational entropy. The former are defined as in Eq.\ (\ref{Hrel}), where the reference state with $\mathbf{u}_i=0$ corresponds to the ideal (unrelaxed) lattice, but the latter are defined with respect to the equilibrium (relaxed) state, which itself depends on the configuration. Although in the harmonic approximation the two types of force constants are identical, anharmonic effects lead to a rather strong dependence of the force constants on bond lengths.\cite{vdW-RMP} Therefore, we do not expect that the ``unrelaxed'' force constants obtained here are sufficiently accurate for the analysis of vibrational entropies or other sensitive measures of the physical vibrational spectra. They are, however, well-suited for the calculations of the relaxation energies, which is our present purpose. As will be further discussed in Sec.\ \ref{anharmonic}, in Cu-Au and Fe-Pt systems the deviations of the relaxation energy from the harmonic approximation are significant only in strongly relaxing configurations, which are statistically negligible in alloys that are not phase-separating. Even in such configurations the effect on the relaxation energy is not very large and can be taken into account with the help of a simple rescaling correction.

It is well-known that the force constants have to satisfy certain identities following from translational and rotational invariance of a crystal subjected to an arbitrary uniform strain and internal displacements.\cite{Born,vdW-RMP} Since these identities should be satisfied in each configuration of the alloy, each independent term in a cluster expansion for the force constants should satisfy all of them. A possible way of constructing such an expansion was suggested in Ref.\ \onlinecite{Triguero}. On the other hand, force constants corresponding to arbitrary central forces automatically satisfy all the invariance relations.\cite{Born} We therefore used a simple parametrization, in which only central (bond-stretching) forces depend on the configuration, while the non-central force components are configuration-independent. We also limited the range of the force constants to second-nearest neighbors and assumed that the central force for bond $i-j$ depends only on the occupation of sites $i$ and $j$. The resulting model has 9 parameters including 6 central-force constants (3 per coordination sphere: for A-A, B-B, and A-B bonds), 2 non-central force constants for nearest neighbors, and 1 isotropic non-central constant for second-nearest neighbor. We also tested more complicated force constant models and found that, although they improve the fitting of the forces induced by lattice displacements, there is little or no improvement in the prediction of the relaxation energies for the systems we have considered.

There may be different strategies for the fitting of the force constant model. For example, one might evaluate the force constants for each input structure and fit them to a model with configuration dependence. However, for relatively simple structures with a small number of atoms per unit cell, the calculation of force constants requires the use of enlarged supercells, which is computationally inefficient. In order to avoid this complication, we adopted a different approach, facilitated by the fact that our input sets at each composition contain a fairly large number of structures (more than 20). For each input structure we calculate the extra forces appearing when one of the atoms is displaced in one of the three Cartesian directions. For this purpose we employed the standard procedure implemented in the VASP code, in which all such inequivalent displacements are automatically generated. While for any given structure this calculation provides information only about the $\Gamma$-point phonons and is usually insufficient to fix all the force constants, we expect that the inclusion of a sufficient number of input structures of relatively small but different sizes and shapes should provide enough information for the fitting of the configuration-dependent force constant parameters. The magnitude of atomic displacements was taken to be 1\% of the cubic lattice parameter, and displacements of both signs were included to reduce the fitting errors. The fit is then performed by linear regression for the set of equations
\begin{equation}\label{fitFCs}
  \mathbf{F}_i-\mathbf{F}_{i0}=\sum_j\hat A_{ij}\mathbf{u}_j
\end{equation}
where $\mathbf{F}_{i0}$ is the (Kanzaki) force acting on the atom at site $i$ in the unrelaxed state with $\mathbf{u}_j=0$.

\begin{figure*}[hbt]
\begin{center}
\includegraphics[width=0.45\textwidth]{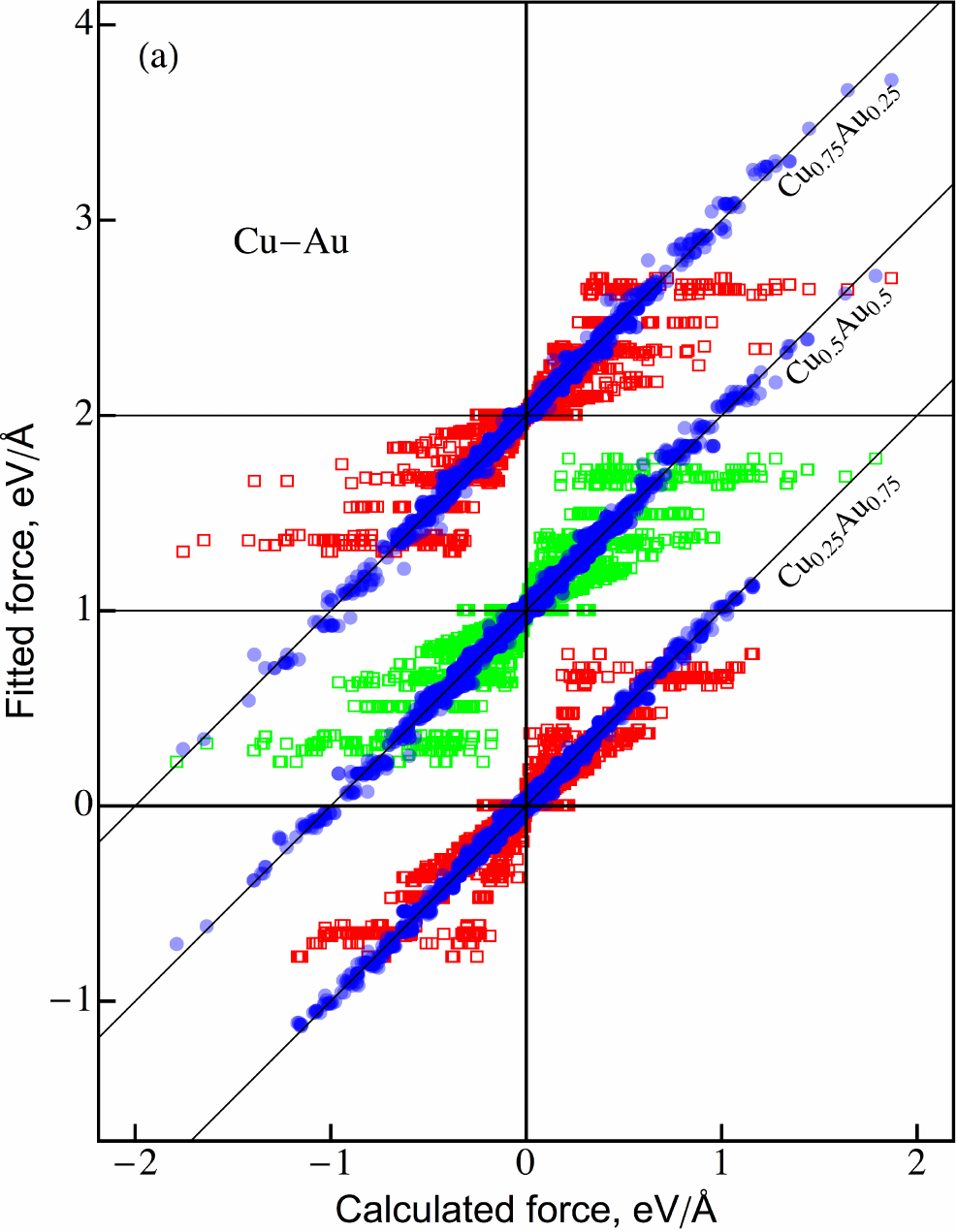}\hfil\includegraphics[width=0.45\textwidth]{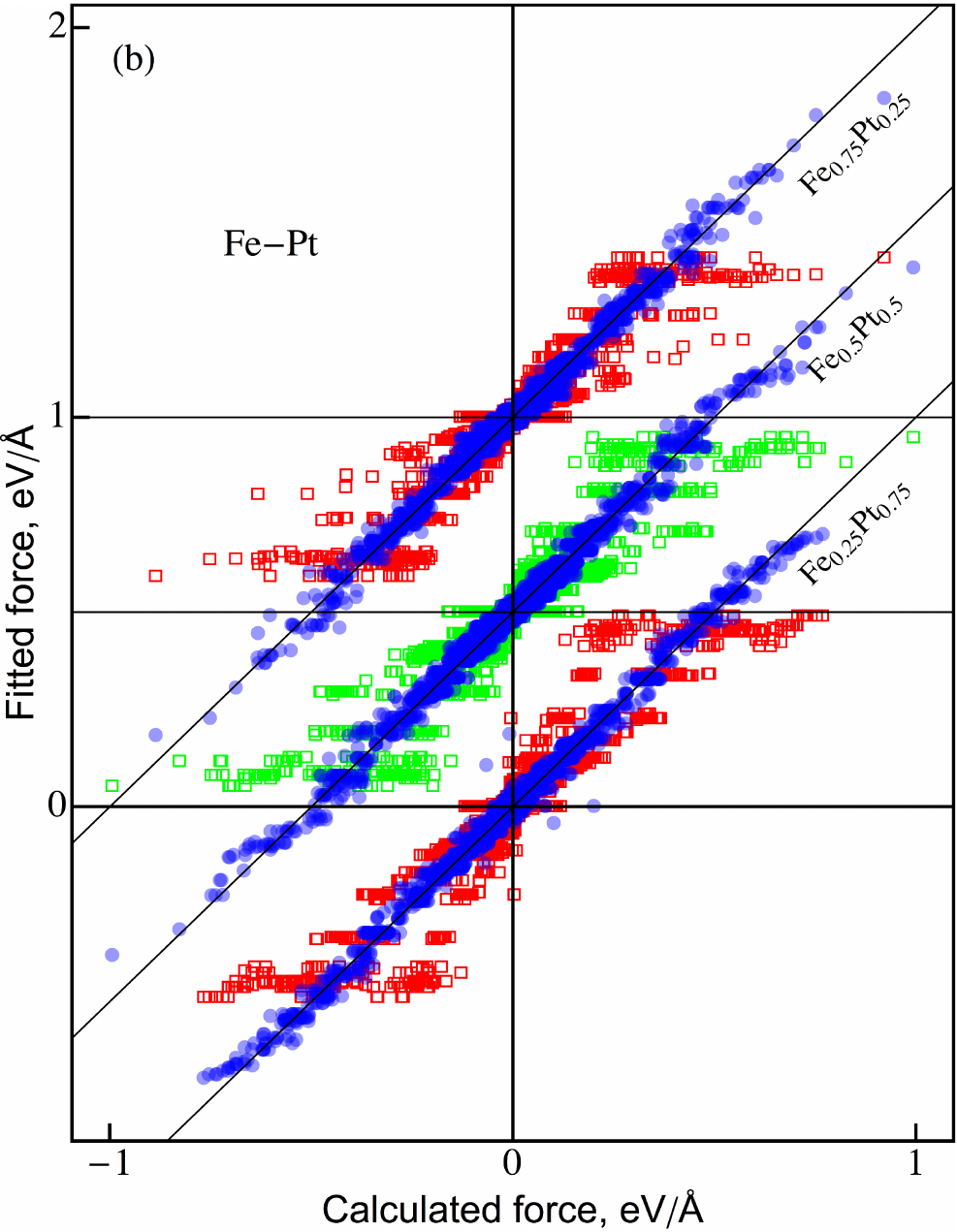}
\end{center}
\caption{(Color online) Fitting of the force constants referenced from the ideal fcc lattice for (a) Cu-Au and (b) Fe-Pt alloys. The forces predicted by the fit are plotted against the calculated forces. Each Cartesian component for each site is represented by one data point. Data sets for different concentrations are shifted along the $y$ axis. Filled (blue) circles and open squares correspond to the models with configuration-dependent and configuration-independent force constants, respectively. See text for details about these models.}
\label{fig:fc}
\end{figure*}

The resulting fit of the configuration-dependent force constants is illustrated in Fig.\ \ref{fig:fc}, along with a similar fit in which the force constants were restricted to be configuration-independent (as it is done in the KKKM). It is seen that the calculated forces are reproduced quite well by a model with configuration-dependent force constants, but the neglect of the configurational dependence results in poor fits for all datasets.

The fitted force-constant parameters are listed in Table \ref{tab:FCs}. Note that the notation for first nearest neighbor force constants is given in a rotated reference frame to isolate the central forces. Specifically, for a (1/2,1/2,0) bond, the axes are rotated by $45^\circ$ around the $z$ axis so that the axis $x'$ lies along the (1,1,0) direction, parallel to the bond. $A_\mathrm{1RR}$ is the central force constant in this rotated frame. One can also write $A_\mathrm{1RR}=(A_\mathrm{1XX}+A_\mathrm{1YY})/2$, $A_\mathrm{1TT}=(A_\mathrm{1XX}-A_\mathrm{1YY})/2$.

Examination of Table \ref{tab:FCs} shows that the nearest-neighbor central force constants are the most important and depend strongly on the identity of the atoms forming the bond. B-B bonds formed by larger atoms of $5d$ elements are very stiff, while A-A bonds formed by smaller $3d$ atoms are relatively soft. This is a common trend in $3d$-$5d$ alloy systems, \cite{vdW-RMP} although we re-emphasize that the force constants used here do not correspond to equilibrium atomic positions.

\begin{table}
\caption{Parameters of the cluster expansion for the force constants (units of dyn/cm). The notation for the first nearest neighbor constants is given in a reference frame with direction $R$ parallel to the bond, $Z$ defined as usual, and $T$ orthogonal to both $R$ and $Z$. 1RR and 2XX generate central forces.}
\begin{tabular}{|c|c|r|r|r|r|r|r|}
\hline
\multicolumn{2}{|c|}{Pairs} & {CuAu$_3$} & {CuAu} & {Cu$_3$Au} & {FePt$_3$} & {FePt} & {Fe$_3$Pt} \\
\hline
\multirow{2}{*}{A-A}
 & 1RR & $10398$ & $13944$ & $18993$ & $10691$ & $18738$ & $28657$ \\
 & 2XX & $14556$ & $9126$ & $5946$ & $20567$ & $13469$ & $4390$ \\
\hline
\multirow{2}{*}{B-B}
 & 1RR & $56877$ & $82571$ & $115172$ & $64987$ & $81150$ & $105670$ \\
 & 2XX & $5599$ & $3172$ & $-3532$ & $10889$ & $10088$ & $3532$ \\
\hline
\multirow{2}{*}{A-B}
 & 1RR & $20930$ & $31434$ & $45081$ & $25229$ & $33785$ & $49155$ \\
 & 2XX & $9282$ & $6529$ & $2989$ & $17976$ & $14457$ & $4729$ \\
\hline
\multirow{3}{*}{Any}
 & 1ZZ & $-11028$ & $-7826$ & $-5026$ & $-6154$ & $-3628$ & $-7257$ \\
 & 1TT & $-3397$ & $-2758$ & $-2150$ & $-2310$ & $-1606$ & $-3387$ \\
 & 2YY & $-1426$ & $-1554$ & $-642$ & $-1966$ & $-3370$ & $-2148$ \\
\hline
\end{tabular}
\label{tab:FCs}
\end{table}

\section{Harmonic relaxation energies} \label{Harmrelaxationen}

\begin{figure}[hbt]
\begin{center}
\includegraphics[width=0.45\textwidth]{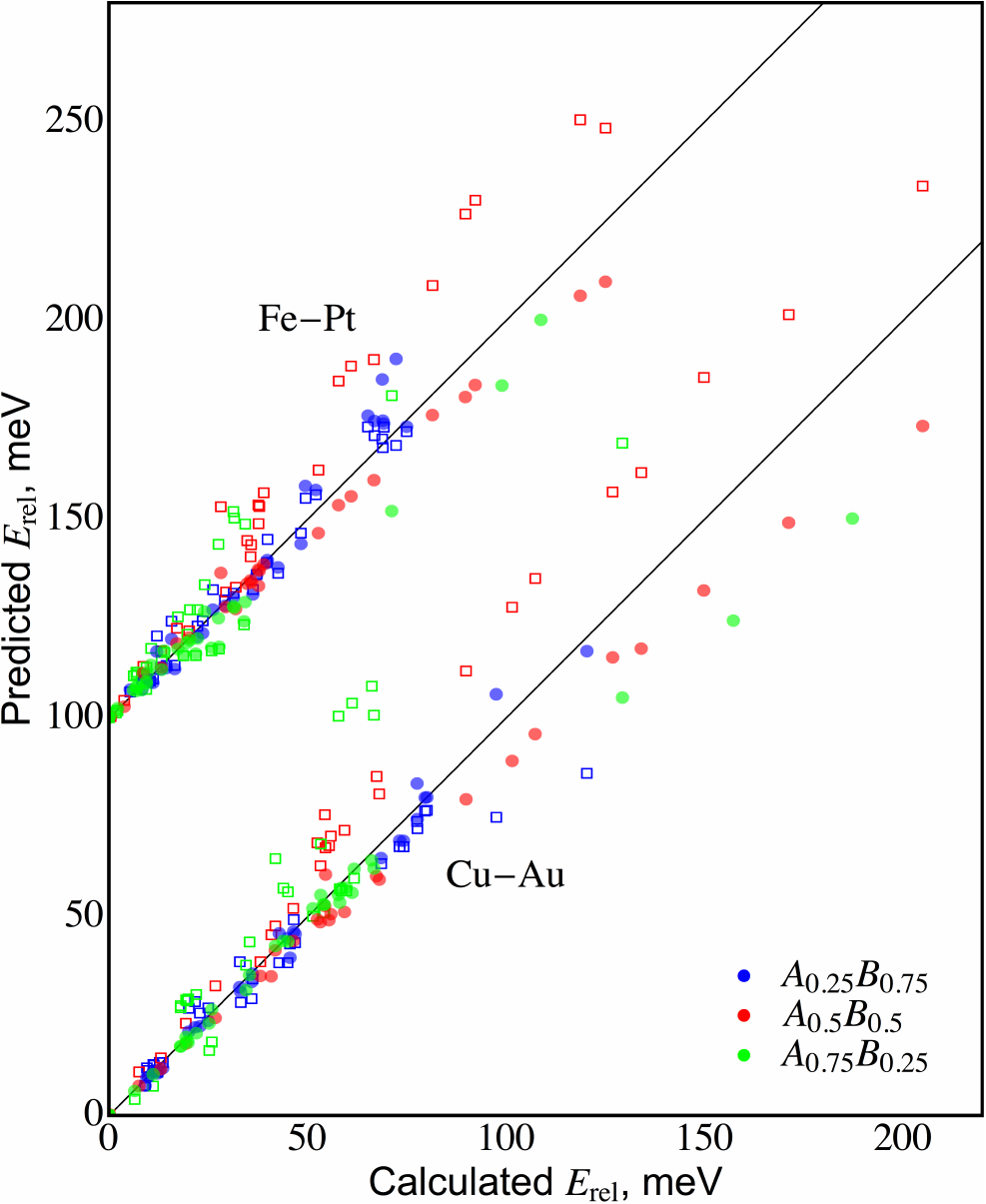}
\end{center}
\caption{(Color online) Relaxation energies predicted using the lattice-deformation model with configuration-dependent (filled circles) or fixed (open squares) force constants. All data for the Fe-Pt system are shifted upwards by 100 meV. Some data points for the model with fixed force constants corresponding to strongly relaxing structures are beyond the field of the figure.}
\label{fig:erel}
\end{figure}

Having constructed the cluster expansions for both Kanzaki forces and force constants, we are now able to calculate the relaxation energies in the harmonic approximation, which are given by Eq.\ (\ref{Erel}). The results of these calculations are illustrated in Fig.\ \ref{fig:erel}, and the corresponding misfits are listed in Table \ref{tab:misfits}. The fidelity of the full configuration-dependent fits (filled circles) is very good, particularly for the structures whose relaxation energy is not very large. Note that the calculated relaxation energies were never used in the construction of this fit. For strongly relaxing structures one can see a systematic underestimation of the relaxation energy, with the points falling below the straight line. This underestimation is due to anharmonicity, as discussed below in Section \ref{anharmonic}. Fig.\ \ref{fig:erel} also shows the predictions of the model in which the force constants are assumed to be configuration-independent (open squares; see also Fig.\ \ref{fig:fc}). It is clear from the figure and from Table \ref{tab:misfits} (first line) that this fit is not very accurate; note that some predicted relaxation energies for strongly relaxing structures fall outside the range of the figure.

One can also see that the fitting of the Kanzaki forces usually does not significantly increase the misfit for the relaxation energies (compare lines 2 and 3 of Table \ref{tab:misfits}). A notable exception is FePt$_3$, where this fitting increases the misfit two-fold.

\begin{table}
\caption{Mean-squared misfits (meV) for the predicted relaxation energies. First two  columns identify the fit as follows. "Exact" forces refers to the calculated forces, while "fitted" forces correspond to the multi-parameter model with the parameters from Table \ref{tab:forces} producing the forces shown by filled circles in Fig.\ \ref{fig:forces}; "fit 2" refers to the 2-parameter model. "Fixed" force constants are configuration-independent (see open squares in Fig.\ \ref{fig:fc}); "full" force constants correspond to the model from Table \ref{tab:FCs} (see also fits shown by filled circles in Fig.\ \ref{fig:fc}).}
\begin{tabular}{|c|c|c|c|c|c|c|c|}
\hline
Forces      & FC          & {CuAu$_3$} & {CuAu} & {Cu$_3$Au} & {FePt$_3$} & {FePt} & {Fe$_3$Pt} \\
\hline
Exact       & Fixed         & 8.0        & 20     & 45         & 4.1        & 19     & 24         \\
Exact       & Full          & 2.7        & 9.0    & 8.9        & 2.4        & 5.2    & 5.1        \\
Fitted      & Full          & 2.8        & 9.3    & 9.7        & 5.7        & 5.5    & 5.8        \\
Fit 2       & Full          & 3.6        & 13.2   & 14.1       & 12.5       & 11.9   & 8.2        \\
\hline
\end{tabular}
\label{tab:misfits}
\end{table}

\section{Role of striction}\label{striction}

The relaxation energies presented in Section \ref{Harmrelaxationen} were, as we mentioned above, calculated under the restriction of zero homogeneous strain. Now we discuss the role of striction for the Cu-Au and Fe-Pt alloy systems.

Denote the relaxation energy at zero strain $E^{sv}_{rel}$ (fixed cell shape and volume), at fixed volume $E^{v}_{rel}$, and $E^{full}_{rel}$ under no restrictions. In the previous sections we were dealing only with $E^{sv}_{rel}$. Now we define $E_{shape}=E^{v}_{rel}-E^{sv}_{rel}$ and $E_{vol}=E^{full}_{rel}-E^{v}_{rel}$ and calculate them for all input structures. The mean values and standard deviations of these energies for each system are listed in Table \ref{tab:VShRel}.

\begin{table}
\caption{Mean values and standard deviations of the relaxation energies corresponding to shape and volume striction.}
\begin{tabular}{|l|c|c|}
\hline
\multirow{2}{*}{Alloy}           & \multicolumn{2}{c|}{Relaxation energy, meV}\\
\cline{2-3}
           & Shape       & Volume \\
\hline
{Fe$_3$Pt} & $38\pm29$   & $1.0\pm1.3$ \\
{FePt}     & $3.3\pm3.0$  	 & $1.3\pm0.8$ \\
{FePt$_3$} & $0.6\pm0.7$ & $1.1\pm0.6$\\
\hline
{Cu$_3$Au} & $9\pm9$     & $9\pm5$\\
{CuAu} (3.819 \AA)     & $14\pm9$    & $11\pm5.5$\\
{CuAu} (3.897 \AA)     & $8.6\pm5.8$    & $3.8\pm2.1$\\
{CuAu$_3$} & $7\pm9$     & $6\pm2$\\
\hline
\end{tabular}
\label{tab:VShRel}
\end{table}

The volume relaxation energies $E_{vol}$ for all three compositions of the Fe-Pt system are small and may be safely neglected. For CuAu$_3$ the standard deviation of $E_{vol}$ is only 2 meV, which means its effect on the phase transitions should be very small. For Cu$_3$Au and CuAu at $a=3.819$ \AA\ the standard deviations of $E_{vol}$ are somewhat larger at 5-6 meV. A more detailed inspection shows that in both cases the variance of $E_{vol}$ is dominated by input structures with total formation enthalpies lying well above the random alloy. For CuAu the mean $E_{vol}$ and its variance are greatly reduced by choosing the lattice parameter by requiring that the volume relaxation energy is minimized on the average ($a=3.897$ \AA). Overall, in all cases the neglect of the volume relaxation energy appears to be a reasonable approximation.

The situation with the shape relaxation energies is more complicated. First, we notice that $E_{shape}$ is very large and has a large variance for the Fe$_3$Pt system. Using a structural filter employed earlier for a similar Fe-Pd alloys,\cite{Chepulskii} we found that a large fraction of Fe$_{0.75}$Pt$_{0.25}$ structures relax to bcc-like final configurations with an accompanying large energy gain. The lowest-energy structures are bcc-like [001] superlattices, suggesting that at low temperatures the system should phase-separate with the precipitation of $\alpha$-Fe. This conclusion appears to be consistent with the experimental phase diagram extrapolated to low temperatures; in this respect the Fe-Pt system is similar to Fe-Pd.\cite{Chepulskii} It may be expected that lattice vibrations stabilize the fcc phase at higher temperatures due to the gain in the vibrational entropy. This stabilization should suppress the tendency toward large shape relaxation for most orderings. Therefore, for the evaluation of the high-temperature ordering phase transition it is reasonable to exclude the shape relaxation in the calculation of the relaxation energies for this system, and this is what we do in the following. For FePt and FePt$_3$ systems the shape relaxation energies are uniformly small and may be neglected.

In the Cu$_3$Au input set there are several structures with $E_{shape}$ exceeding 20 meV, all of which lie above the random alloy; $E_{shape}$ tends to be relatively large for [001] superlattices. Since the L1$_2$ ground state and the random alloy have no shape relaxation, one can expect that the shape relaxation energy has a small effect on the ordering phase transition. However, more delicate properties such as short-range may require $E_{shape}$ to be included.

For the CuAu composition the shape relaxation is very important, because it stabilizes the L1$_0$ ground state which has a large tetragonal distortion with the theoretical $c/a$ ratio as low as 0.925 (compare to 0.963 for FePt).
The corresponding $E_{shape}$ is 17.5 meV at $a=3.819$ \AA\ and 12.4 meV at $a=3.897$ \AA. Strong shape relaxation of [001] superlattices is thus shared with the Cu$_3$Au system.
We found that if $E_{shape}$ is not included, at least one 16-atom structure comes within 2-3 meV of the formation enthalpy of L1$_0$ and tends to appear in Monte Carlo simulations due to a small error in the refitted cluster expansion (see below). This structure is shown in Fig.\ \ref{8888}. Its shape relaxation energy is less than 1 meV, and the inclusion of $E_{shape}$ therefore strongly favors the L1$_0$ ordering; it is clear that this must be taken into account in thermodynamic calculations (see Section \ref{mcresults}).

\begin{figure}[hbt]
\centering
\includegraphics[width=0.3\textwidth]{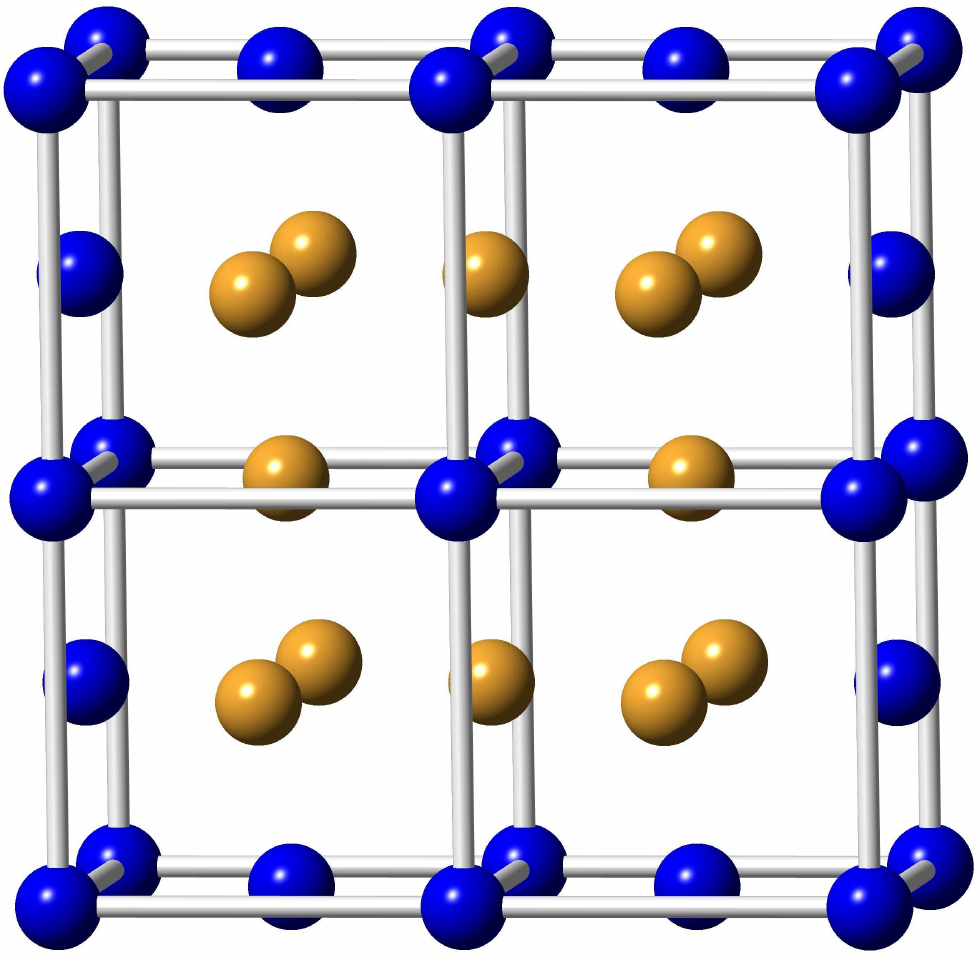}
\caption{(Color online) 16-atom structure X competing with L1$_0$ in the CuAu system.}
\label{8888}
\end{figure}

In the CuAu$_3$ input set there are three structures with $E_{shape}$ of about 30 meV accounting for most of the variance, which are all [001] superlattices. Their fully relaxed formation enthalpies are nearly degenerate, lie at the lower end of the spectrum of all the calculated structures, which is about 10 meV lower than the L1$_2$ structure. This is a reproduction of the known result that GGA predicts an incorrect ground state for CuAu$_3$.\cite{Ozolins2} Moreover, these lowest-energy structures are also nearly degenerate with two other structures whose shape relaxation energies are only about 4 meV, and which, therefore, remain lower than L1$_2$ even if $E_{shape}$ were neglected. Thus, even if we assume that the shape relaxation energy is suppressed at higher temperatures and should be neglected, the L1$_2$ phase remains unfavorable based on its formation enthalpy. The results reported in Section \ref{mcresults} show that this phase indeed does not appear at elevated temperatures under this assumption.

In principle, the uniform strain contribution could be evaluated by minimizing $H_{rel}$ with respect to both $\mathbf{w}_i$ and $u_{\alpha\beta}$. We found, however, that the stress tensor
$\sum\limits_{ij}F^\alpha_{ij}(\Sigma)R^\beta_{ij}$ calculated at $\mathbf{w}_i=0$ in Eq.\ (\ref{Hrel5}) is not well predicted using the fitted Kanzaki forces in any of the studied input sets. For Cu-Au systems the misprediction could be largely eliminated by multiplying the stress tensor by an overall factor, but for the Fe-Pt systems the prediction barely correlates with the calculated stress tensor. It is likely that the calculation of the stress tensor converges slowly in real space, even though for the forces it is sufficient to include interactions from only a few coordination spheres. In order to study the convergence of the stress tensor in real space it would be necessary to design an automatic procedure to construct a complete basis set for the cluster-expanded Kanzaki forces. Leaving this problem for future studies, here we resort to a conventional cluster expansion to represent the uniform strain contribution in the Cu$_{0.5}$Au$_{0.5}$ system as explained in Section \ref{mcresults}.

\section{Anharmonicity} \label{anharmonic}

In Section \ref{Harmrelaxationen} we saw that Eq.\ (\ref{Erel}) systematically underestimates the relaxation energy for structures where it is large. This inaccuracy could be due to an imperfect representation of the forces and force constants in the model and due to anharmonicity. For some insight into this issue, we focussed on the Cu$_{0.5}$Au$_{0.5}$ system. We picked 8 input structures including some with relatively large relaxation energies, and added the structure that was predicted by Monte Carlo calculations based on CLDM to be the ground state if uniform strain is neglected (see below). For each of these structures we took the calculated equilibrium atomic displacements $\mathbf{w}^{eq}_i$, defined a continuous path $\mathbf{w}_i(t)=t \mathbf{w}^{eq}_i$, and calculated the relaxation energy $E_{rel}$ as a function of $t\in[0,1]$. For all structures $E_{rel}$ can be well approximated by a third-order polynomial
\begin{equation}
E_{rel}=-ft+\frac12at^2+\frac13bt^3
\label{unharmen}
\end{equation}
The parameters $f$ and $a$ can be directly compared with the predictions of the CLDM fits. Further, in our harmonic CLDM with $b=0$ the equilibrium value of $t$ is $t_0=f/a$, and the relaxation energy is $E^{harm}_{rel}=-f^2/(2a)$. (In this crude treatment we disregard the fact that the relaxation path itself depends on the choice of the model.) The results of this analysis are listed in Table \ref{tab:anharmstruc}.

\begin{table}[hbt]
\caption{Parameters of Eq.\ (\ref{unharmen}) fitted to VASP calculations and relaxation energies compared to CLDM predictions (see text for details). All values are in meV.}
\begin{tabular}{|c|c|c|c|c|c|c|}
 \hline
Structure      &  Source  & $f$  & $a$ & $-b$ &$E^{harm}_{rel}$ & $E_{rel}$\\
\hline
\multirow{2}{*}{A$_2$B$_2$ [001]}       	& VASP  &261.1       & 179.9       & 33.9    &94.7 &115.0 \\
						& CLDM  &262.8       & 185.6       &  0   &93.0 &\\
\hline
\multirow{2}{*}{A$_2$B$_2$ [011]}     & VASP& 110.3 &56.0     &0.9 &54.3&55.2\\
	                & CLDM& 105.6 &59.9 &0    &46.6&\\
\hline
\multirow{2}{*}{A$_2$B$_2$ [111]}       & VASP    & 242.5       & 163.1       &30    &90.1 & 109.1\\
					& CLDM    & 243.3       & 175.0    &  0  &84.6 &\\
 \hline
\multirow{2}{*}{A$_3$B$_3$ [001]} & VASP    &347.9       &232.3       & 40.3    &130.2& 155.6\\
				     & CLDM    &345.8       &237.7       &  0   &125.8&\\
\hline
\multirow{2}{*}{A$_3$B$_3$ [111]} & VASP    & 293.0       & 191.6       & 32.0    &112.0& 133.2\\
				     & CLDM    & 292.9       & 198.7       &  0   &107.9&\\
\hline
\multirow{2}{*}{A$_2$B$_2$AB [001]} & VASP    & 179.1       & 125.6       & 24.9   &63.8& 78.2\\
					 & CLDM    & 179.4       & 130.1       & 0   &61.8 &\\
 \hline
\multirow{2}{*}{A$_2$B$_2$AB [111]} & VASP    &167.3       & 120.6       &25.9    & 58.0 & 72.4\\
					  & CLDM    &168.2       & 128.1       & 0   &55.3 &\\
 \hline
\multirow{2}{*}{A$_2$B$_2$AB [133]} & VASP    &112.8 & 69.6       & 8.9      & 45.8    &52.1\\
				         & CLDM    &113.5 & 70.2       &  0    & 45.9    &\\
 \hline
\multirow{2}{*}{X}         & VASP   &86.0    & 44.9       & 1.3       & 41.1    &42.3\\
	& CLDM   &89.7    & 43.0       &   0     & 46.8  &\\
\hline
\end{tabular}
\label{tab:anharmstruc}
\end{table}

As can be seen from Table \ref{tab:anharmstruc}, the ``force'' and ``stiffness'' parameters $f$ and $a$ given by the CLDM fits agree quite well with those obtained from VASP,  and the relaxation energies predicted by CLDM are close to the harmonic approximation $E^{harm}_{rel}$ based on VASP results. The largest discrepancy of nearly 8 meV occurs for the A$_2$B$_2$ [011] structure, where $f$ is underpredicted by 4.3\% and $a$ is overpredicted by 7\%. On the other hand, the error introduced by neglecting the anharmonicity reaches 20-25 meV for strongly relaxing structures. For all structures shown in Table  \ref{tab:anharmstruc} the parameter $b$ is negative, which leads to an underestimated relaxation energy in the harmonic approximation. Negative $b$ means that the force constants decrease (i.\ e.\ bonds soften) in the course of the relaxation. Note that the actual predictions of CLDM are closer to the VASP values compared to those found in Table \ref{tab:anharmstruc}, because the relaxation path in CLDM differs from the one in the full calculation.

These results strongly suggest that the underestimation of the relaxation energies in our implementation of CLDM is primarily due to its reliance on the harmonic approximation. Although the accuracy of its predictions for Cu-Au and Fe-Pt systems is acceptable (particularly because the largest discrepancies occur in strongly relaxing structures that tend to have large total formation enthalpies and are therefore statistically insignificant), we see that further improvement may be achieved by going beyond the harmonic approximation. We leave this problem for future studies.

\section{Phonon spectra}\label{sec:phonons}

The configuration-dependent representation of the force constants makes it possible to evaluate the phonon spectra for any given configurational state of the alloy. The spectrum of the stable ordered structure and that of the random alloy can be compared with experiment and are therefore of primary interest. One property of great importance that is governed by the phonon spectrum is the vibrational entropy, which contributes to thermodynamics and modifies the transition temperatures. \cite{vdW-RMP}

In the harmonic approximation the force constants do not depend on the reference state, be it the ideal lattice or equilibrium positions. In reality, however, anharmonicity makes the force constants depend rather strongly on bond lengths in alloys with large size mismatch. \cite{vdW-RMP} Therefore, within the harmonic model we can only obtain a rough approximation for the phonon spectrum of the random alloy. In particular, we found that this approximation is not suitable for the evaluation of the vibrational entropy: The force constants for the random alloy are too large, and the vibrational entropy change under ordering often comes out positive in contradiction with earlier results. \cite{vdW-RMP} An anharmonic extension of the CLDM can provide a reasonable approach to such calculations. (See Ref.\ \onlinecite{TDEP} for an application of a similar model to finite-temperature lattice dynamics in pure compounds.) Nevertheless, we expect that the general effects of alloy disorder on the phonon spectra should be reasonably well captured even within the harmonic approximation.

\begin{figure*}[hbt]
\includegraphics[width=0.95\textwidth]{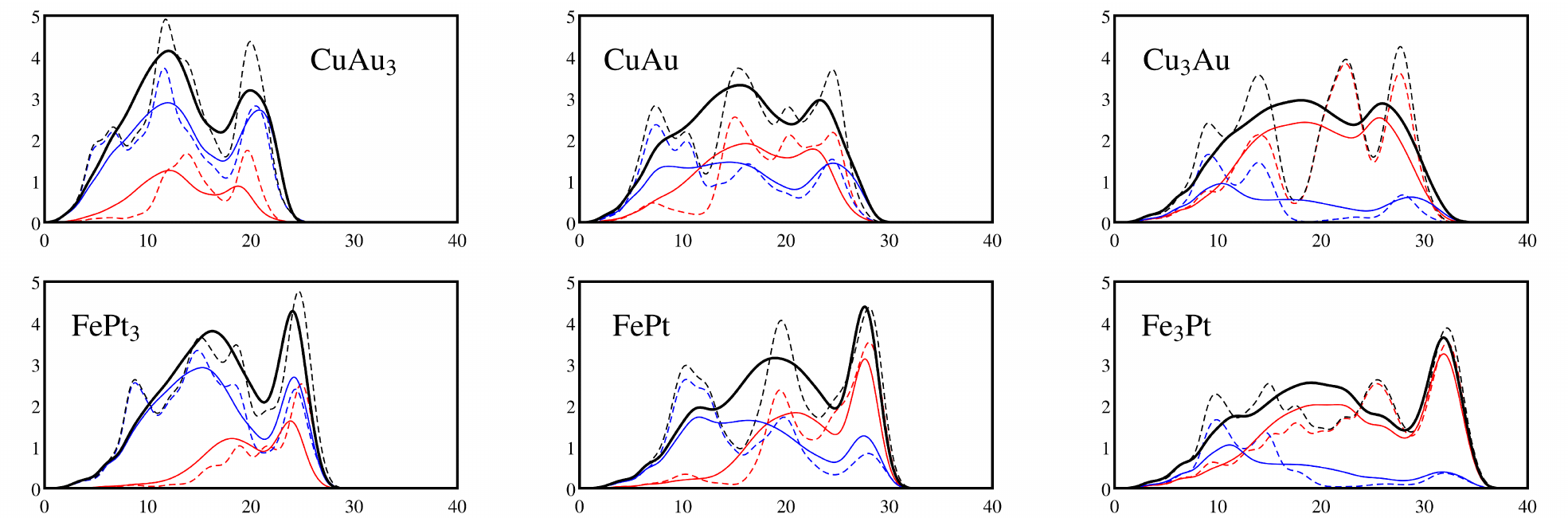}\hfil
\caption{(Color online) Phonon spectra calculated using a $8\times8\times8$ cubic cell containing 2048 atoms and the fitted force constants. Solid lines: random alloy. Dashed lines: fully ordered alloy (L1$_0$ for 50\% and L1$_2$ for 1:3 compositions). Black lines: total phonon DOS; blue and red lines: partial DOS for $5d$ and $3d$ elements, respectively. The frequency scale is in meV; DOS is in arbitrary units.}
\label{fig:phonons}
\end{figure*}

In order to calculate the phonon density of states (DOS), we construct a cubic supercell with the given configurational state of the alloy (either fully ordered L1$_0$ or L1$_2$ or disordered), compute the phonon frequencies for this supercell assuming periodic boundary conditions, and plot the DOS using Gaussian smearing. The phonon DOS is practically identical for different realizations of the random alloy in a cubic 2048-atom supercell. Apart from the contribution of the low-frequency acoustic phonons, this calculation therefore provides a very good representation of the phonon DOS. In addition to the total DOS, we calculated the partial contributions of the two alloy components using the eigenvectors of the phonon modes. The results are shown in Fig.\ \ref{fig:phonons}.

The phonon spectrum for both ordered and disordered Cu$_3$Au are quite similar to experimental data. \cite{Bogdanoff-03} The positions of the peaks for the L1$_2$ structure at 9, 13, 22 and 28 meV are in good agreement with the experimental positions at 9, 13, 21, and 25 meV. In agreement with Ref.\ \cite{Bogdanoff-03}, the 9 meV peak is a resonance dominated by the motion of Au atoms, as can be seen from the partial DOS. The peak at 13 meV has a slighly larger total Cu weight, which translates to the Au atoms oscillating with about twice as large an amplitude as the Cu atoms. The peak at 22 meV corresponds to an almost pure Cu oscillation, and the peak at 28 meV has similar amplitudes for all atoms. In the disordered alloy we observe a strong smearing of all peaks, while the high-energy peak moves down by about 2 meV. The Au resonance is still seen, but it is weakened compared to the ordered alloy and moved a little higher to 10 meV. The peaks at 13 and 22 meV are smeared out into one broad maximum with a substantial Au spectral weight. The latter was interpreted as the result of the removal of constraints on the motions of Cu atoms by the framework of the heavy Au atoms due to disorder. \cite{Bogdanoff-03}

In disordered CuAu we also observe an Au resonance at 8-9 meV producing a shoulder in the spectrum in good agreement with experiment. \cite{Bogdanoff-03} The higher-energy peaks at 15 and 23 meV are also in good agreement with the experimental positions of 15 and 22 meV. \cite{Bogdanoff-03}

In CuAu$_3$ the Au resonance broadens and only a weak shoulder remains. The peaks at 12 and 20 meV can be matched with the broad shoulder at 12-13 meV and a peak at 18-19 meV in experiment. \cite{Bogdanoff-03} Thus, all the spectral features for Cu-Au alloys are in good agreement with experiment, except that the positions of the high-frequency peaks are about 1 meV too high.

The phonon spectrum of FePt$_3$ shows a strongly Fe-weighted peak at 25 meV in excellent agreement with experimental data. \cite{Yue,Noda} The spectum of the ordered L1$_2$ FePt$_3$ agrees well with Ref.\ \onlinecite{Noda}. In the disordered alloy the peaks broaden, and the highest-enegry peak shifting slightly down.

Overall, the spectra for the Fe-Pt alloys are similar to Cu-Au, but the highest frequencies are higher, and the corresponding peaks sharper. This is due to Fe having a smaller mass and the Fe-Fe and Fe-Pt bonds being stiffer compared to Cu-Cu and Cu-Au.
The spectrum for Fe$_3$Pt reveals a Pt resonance at 10 meV, which is similar to the Au resonance in Cu$_3$Au.

\section{Auxiliary cluster expansion}\label{auxce}

In thermodynamic or kinetic Monte Carlo simulations the formation enthalpy of a simulation cell with at least a few thousand atoms has to be calculated multiple times. The chemical part $H_{chem}$ (see Eq.\ \ref{Htot}) can be represented by a conventional real-space cluster expansion. Unfortunately, the CLDM expressions for the relaxation energy can not be employed directly in Monte Carlo simulations due to a prohibitive cost of inverting, at each Monte Carlo step, the matrix $\hat A$ in Eq.\ (\ref{Erel}) whose dimension scales with the number of atoms in the simulation cell. On the other hand, for a qualitative analysis of ordering tendencies and phase transformation kinetics, it is often useful to analyze the energetics of an alloy that is close to random using the concentration-wave method. \cite{Krivoglaz,Khachaturyan} While more sophisticated approximations can be envisaged, here we address both of these needs by fitting the CLDM relaxation energy to an auxiliary many-body real-space cluster expansion. Such an expansion can have a much larger number of parameters compared to those based directly on first-principles data, because a large number of structures can be included in the fit at a small computational cost. Because of this, the problem of convergence in real space is greatly alleviated. The price to pay for this simplification is, of course, the loss of information about the long-range part of the interaction. This loss is likely not critical for systems undergoing homogeneous ordering, but it is troublesome whenever phase separation is involved. Indeed, the long-range part of the interaction is responsible for the non-additive contribution to the relaxation energy which is crucial in coherent phase separation.\cite{Williams1,Williams2,CL,Larche,Johnson} Since our present focus is on ordering systems, we expect the auxiliary cluster expansion for the CLDM energy to be a good approximation.

\begin{figure*}[hbt]
\centering
\includegraphics[width=0.4\textwidth]{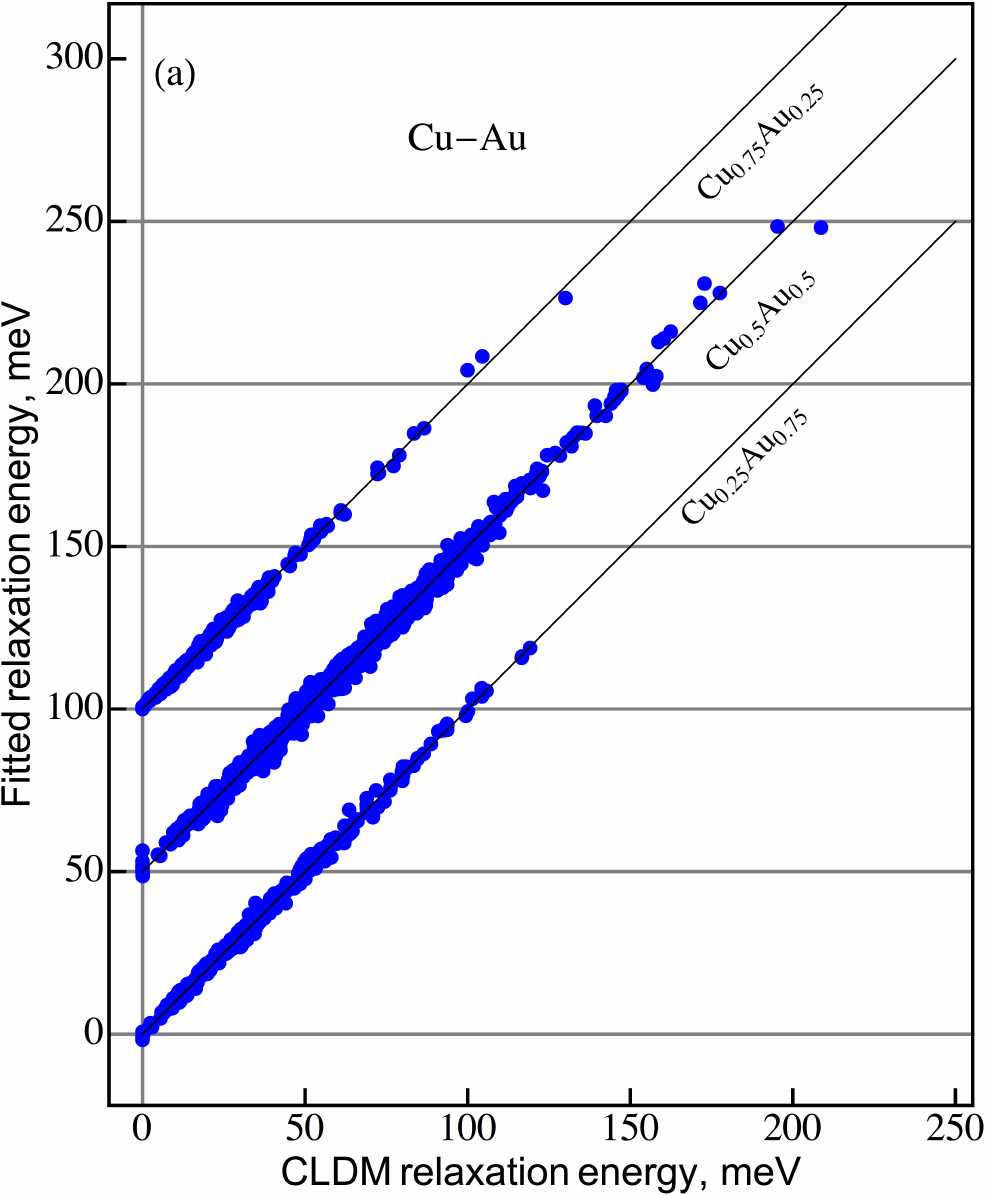}\hfil\includegraphics[width=0.4\textwidth]{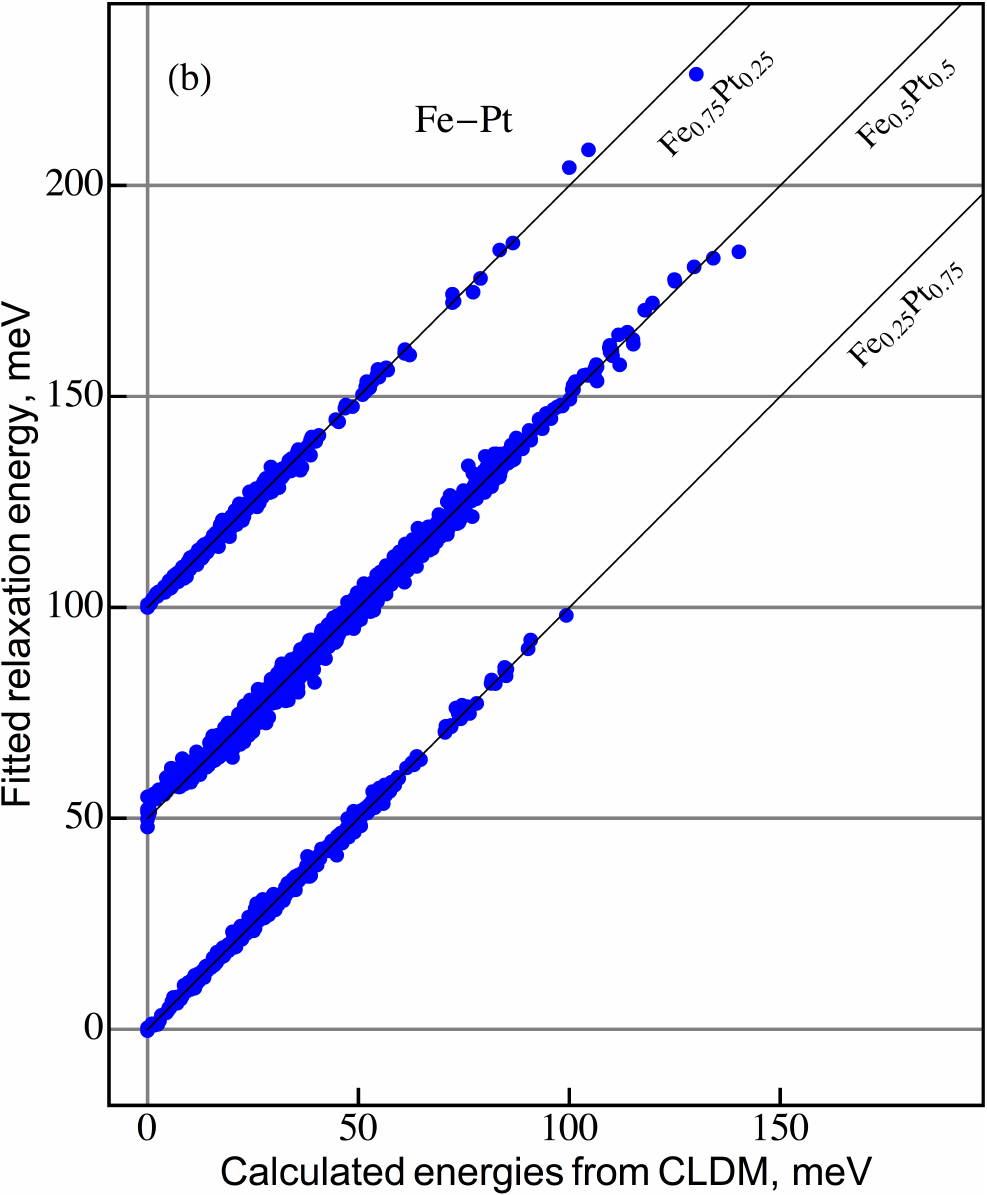}
\caption{(Color online) Accuracy of the real-space cluster expansions of the CLDM relaxation energy for (a) Cu-Au and (b) Fe-Pt alloys. Each structure is represented by a data point. Data sets for different concentrations are shifted along the y axis.}
\label{fig:atatexp}
\end{figure*}

We have calculated the CLDM relaxation energies for a few hundred structures at each target composition for Cu-Au and Fe-Pt systems, and used them to construct real-space cluster expansions for each of the three concentrations by means of the Alloy Theoretic Automated Toolkit (ATAT).\cite{Walle} All input structures were generated at the exact target concentration. The basis sets and overall quality of these expansions are displayed in Table \ref{tab:clexp} (see the lines labeled R). A more detailed picture is provided by Fig.\ \ref{fig:atatexp}. It can be seen that the misfits introduced by the re-expansion of the CLDM relaxation energy are smaller than 3 meV. In addition, the CV scores are only marginally larger than the average misfits, reflecting the stability of the fits thanks to the fact that there are many more input structures than fitting parameters. Note that these misfits reflect only the additional error introduced by re-expanding the CLDM expressions; the accuracy of the CLDM itself was discussed in Section \ref{Harmrelaxationen}.

\begin{table}[hbt]
\caption{Parameters of the real-space cluster expansions used in Monte Carlo simulations. The column \emph{Clusters} lists the number of 2-body, 3-body, 4-body, and larger (for CuAu) cluster types in the basis set. The CV and misfit are given in meV. Notation: GS, tentative ground state predicted by the full cluster expansion; OV, at optimized volume; R, relaxation energy from CLDM; C, chemical part of the formation enthalpy; S, contribution from striction; NM, not meaningful; [001] SL, several [001] superlattices are nearly degenerate. Structure Y for FePt$_3$ is a [001] A$_2$BA$_2$M superlattice, where M is a mixed A/B layer; it is nearly degenerate with L1$_2$.}
\begin{tabular}{|l|c|c|c|c|c|c|}
\hline
\multicolumn{1}{|c|}{Alloy} & Term     & Inputs & Clusters & CV & Misfit & GS\\
\hline
\multirow{2}{*}{Fe$_3$Pt} & R   & 440   & 39,50,20       & 1.8   & 1.0  &\multirow{2}{*} {L1$_2$} \\
\cline{2-6}
			              & C   & 33	   & 4		           & 3.6   & 3.2       &\\
\hline
\multirow{2}{*}{FePt}     & R  	& 1936   & 39,50,20        & 1.8  & 1.6     &\multirow{2}{*} {L1$_0$}   \\
\cline{2-6}
			              & C   & 28	        & 4,0,1		& 6.0    &  4.7      &\\
\hline
\multirow{2}{*}{FePt$_3$} & R   & 440   & 39,50,20       & 1.4   & 1.0   &\multirow{2}{*} {Y}    \\
\cline{2-6}
			              & C   &  33	      & 3,2		     & 3.8   &  3.1      &\\
\hline
\hline
\multirow{2}{*}{Cu$_3$Au} & R   & 439   & 39,50,35       		& 2.8 &    1.0    & \multirow{2}{*}{L1$_2$}   \\
\cline{2-6}
			              & C   & 33	      & 4		     &1.5    &  0.6      &\\
\hline
\multirow{3}{*}{CuAu}     & R  	& 1195   & 39,50,30,17,20        & 2.4      &  2.1     & \multirow{3}{*}{L1$_0$}    \\
\cline{2-6}
			              & C   & 29	      & 7		     &1.3    &  1.0      &\\
\cline{2-6}
			              & S   & 26	      & 7,1,5	     & NM   &  2.9      &\\
\hline
\multirow{3}{3em}{CuAu (OV)}  & R     	& 1195  & 39,50,30,17,20        & 2.0      &  1.7     & \multirow{3}{*}{L1$_0$}    \\
\cline{2-6}
			                  & C       & 29	      & 6		     &1.2    &  0.9      &\\
\cline{2-6}
			                  & S       & 47	      & 7,1,5	     & 6.1   &   3.3     &\\
\hline
\multirow{2}{*}{CuAu$_3$} & R  & 439   & 39,50,35       	        & 2.0  & 1.3      & \multirow{2}{*}{[001] SL}     \\
\cline{2-6}
			              & C  & 33	      & 4		     &1.8    &1.5        &\\
\hline
\end{tabular}
\label{tab:clexp}
\end{table}

\begin{figure*}[hbt]
\centering
\includegraphics[width=0.45\textwidth]{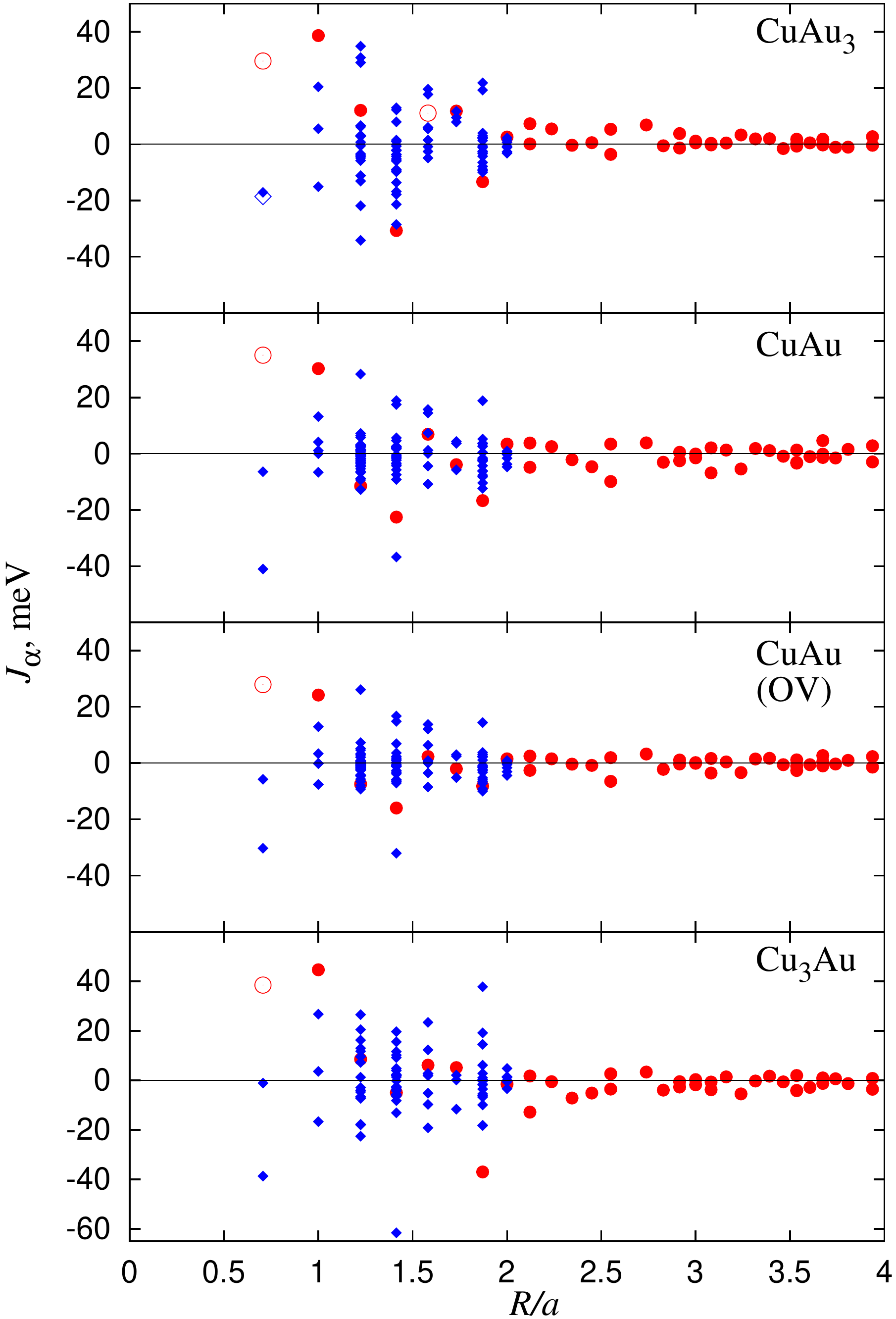}\hfil\includegraphics[width=0.45\textwidth]{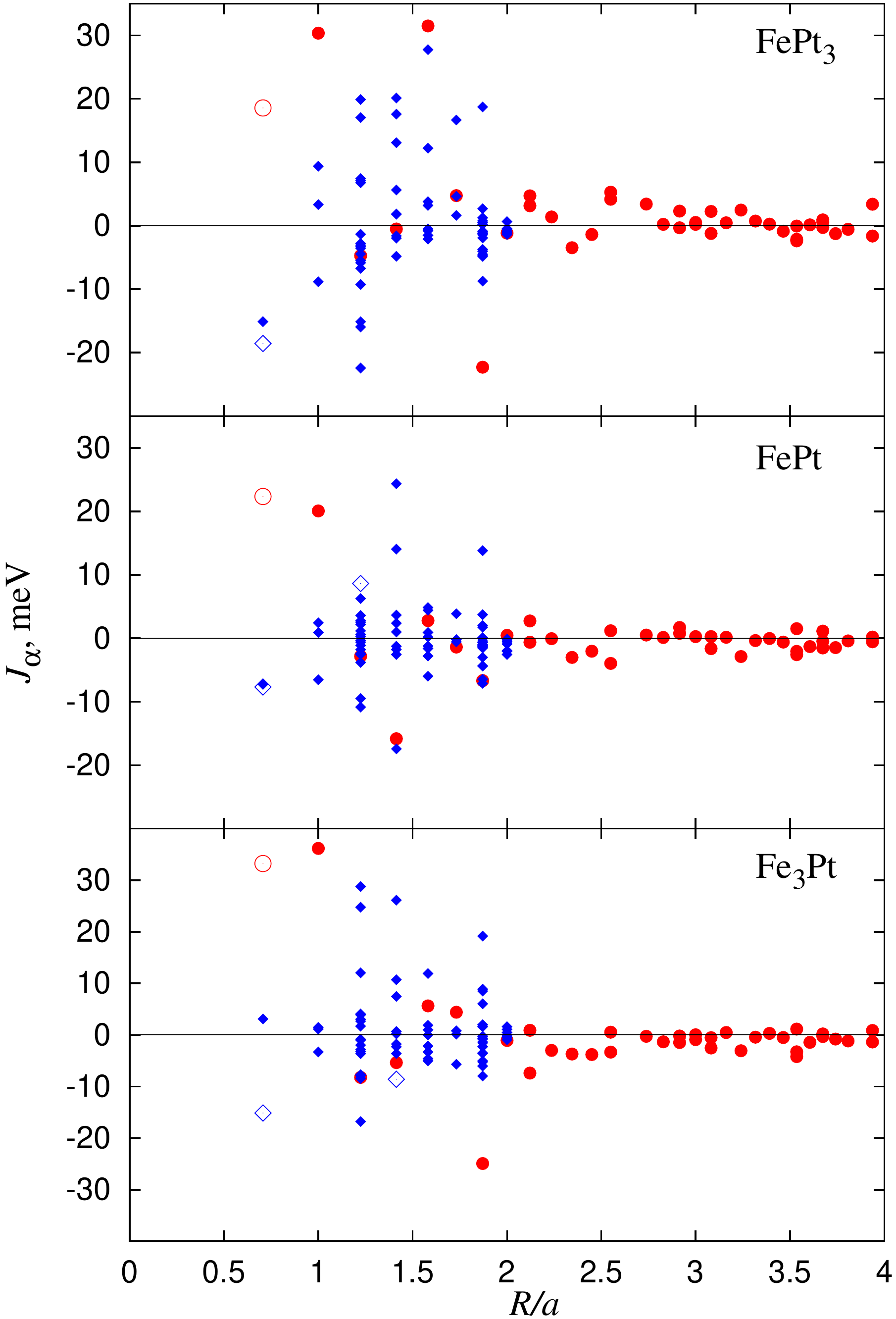}
\caption{(Color online) Effective cluster interactions (ECI) representing the strain-induced interaction in Cu-Au (left panel) and Fe-Pt (right panel) alloys. Red circles: pair ECI; blue diamonds: many-body ECI. Large empty symbols: these ECIs were divided by 5 to fit in the figure. The ECIs are multiplied by the multiplicity factor giving the number of the corresponding clusters per lattice site. OV: optimized volume.}
\label{fig:ECI}
\end{figure*}

Fig.\ \ref{fig:ECI} illustrates the dependence of the effective cluster interactions (ECI) representing the CLDM relaxation energy on the size of the cluster (i.\ e.\ largest distance between any two sites within the cluster). Two-body and many-body ECIs are shown separately, and they are multiplied by the factor giving the number of clusters of a given type per lattice site. It is seen that the many-body ECIs are comparable with pair ECIs. Moreover, although many-body clusters with size larger than $2a$ were not included in the basis sets, Fig.\ \ref{fig:ECI} strongly suggests that many-body ECIs, similarly to the pair ones, decay slowly with the cluster size. This feature agrees with expectations based on the general structure of CLDM and the fact that the Kanzaki forces are strongly non-pairwise.

The ``chemical'' (fixed lattice) contribution to the formation enthalpy is represented by a conventional real-space cluster expansions, whose parameters are listed in Table \ref{tab:clexp} (lines labeled C). The inputs for these expansions were the same as those used in the fitting of the forces and force constants. The Cu-Au set also included structure X shown in Fig.\ \ref{8888} which competes with L1$_0$ if the striction term is not included.
As noted in Section \ref{striction}, this striction is important for the stabilization of the L1$_0$ ground state and must be included in thermodynamic simulations. Since our representation of the forces and force constants does not provide an adequate mapping of the stress tensor within CLDM, for this system we utilized an additional real-space cluster expansion to represent the striction part of the formation enthalpy $E_{shape}+E_{vol}$ (in the notation of Section \ref{striction}). The data for the construction of these cluster expansions is taken from first-principles calculations with fixed and optimized dimensions of the unit cell. The parameters of this expansion are also included in Table \ref{tab:clexp}. The structure shown in Fig.\ \ref{8888} is not invariant under relabeling of the component species, which introduces a three-body term in the cluster expansion. If this is the only such structure in the input set, the CV score becomes meaningless. Therefore, in the fit at the optimized volume we have included several additional structures. Based on the discussion in Section \ref{striction}, the contribution of striction was neglected for the other five systems considered here.

\section{Effective interaction in the random alloy}\label{sec:veff}

Let us now consider the effective pair interaction that may be suitable for the analysis of the stability of the random alloy with respect to the formation of concentration waves. \cite{Krivoglaz,Khachaturyan} While short-range order may complicate the situation considerably, here we restrict ourselves to the consideration of an ensemble with statistically independent atomic site occupations representing alloy configurations that are only slightly inhomogeneous. This means that the average occupations are $\langle\sigma_i\rangle=\sigma_0 + \delta_i$ where $\delta_i$ are small.
The effective interaction potential is then defined as:
\begin{equation}
J^\mathrm{eff}_{ij} = \frac{\partial^2 \langle E\rangle}{\partial \delta_i\partial \delta_j} .
\end{equation}
If the energy is represented by a many-body real-space cluster expansion, the effective potential can be readily calculated:
\begin{equation}
J^\mathrm{eff}_{ij} = \sum_{P\supset \{i,j\}}  \frac{m_P N^P_{ij}J_P\sigma_0^{N_P-2}}{m_{ij}}\label{jeff}
\end{equation}
where $m_P$ is the multiplicity factor of cluster type $P$, $N_P$ is the number of sites in $P$, and $N^P_{ij}$ the number of edges of $P$ that are equivalent by symmetry to the pair $\{i,j\}$. Note that for an equiconcentrational alloy we have $\sigma_0=0$, and only pair clusters contribute to $J^\mathrm{eff}_{ij}$.

\begin{figure*}[hbt]
\centering
\includegraphics[width=0.45\textwidth]{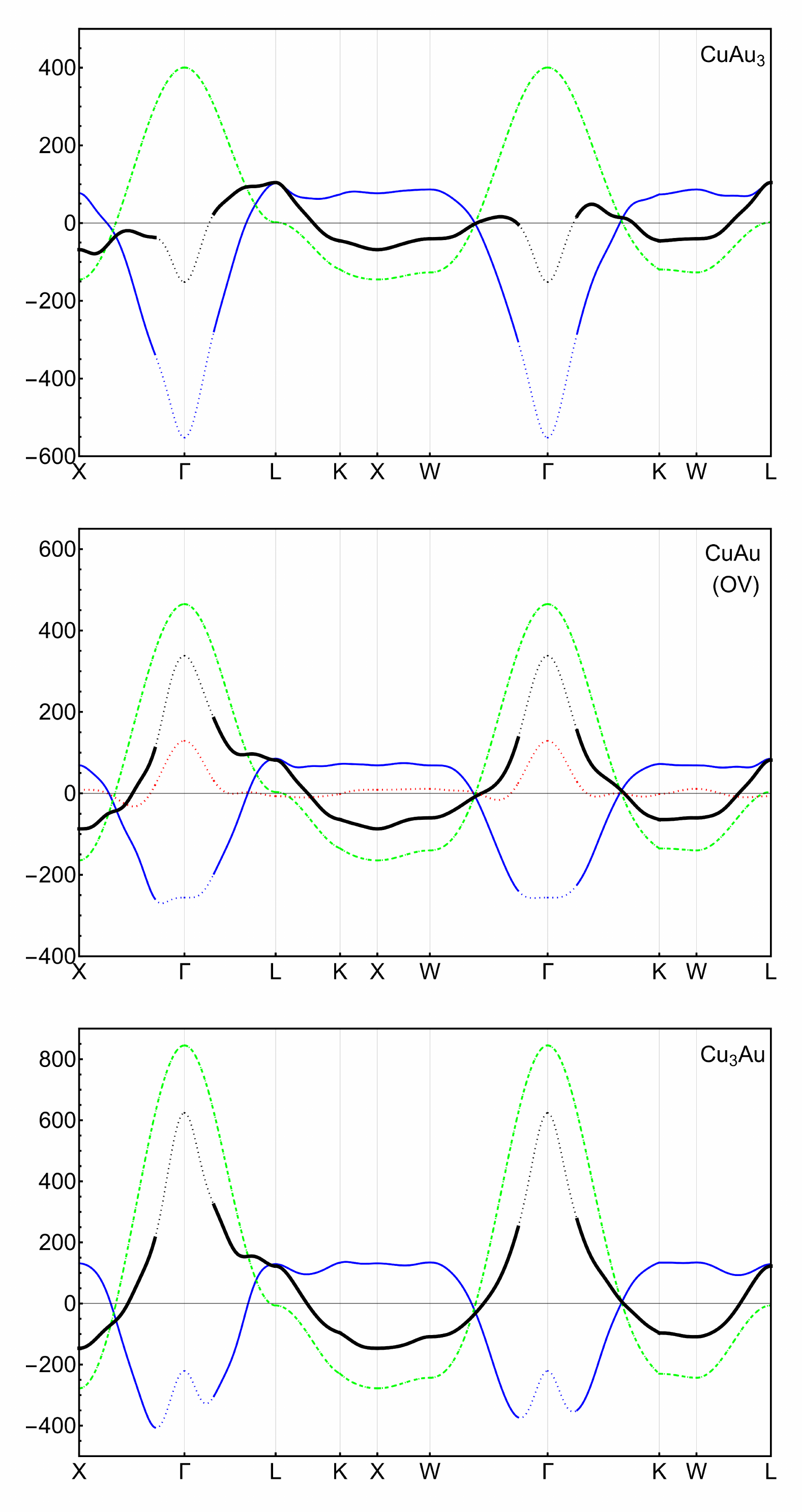}\hfil\includegraphics[width=0.45\textwidth]{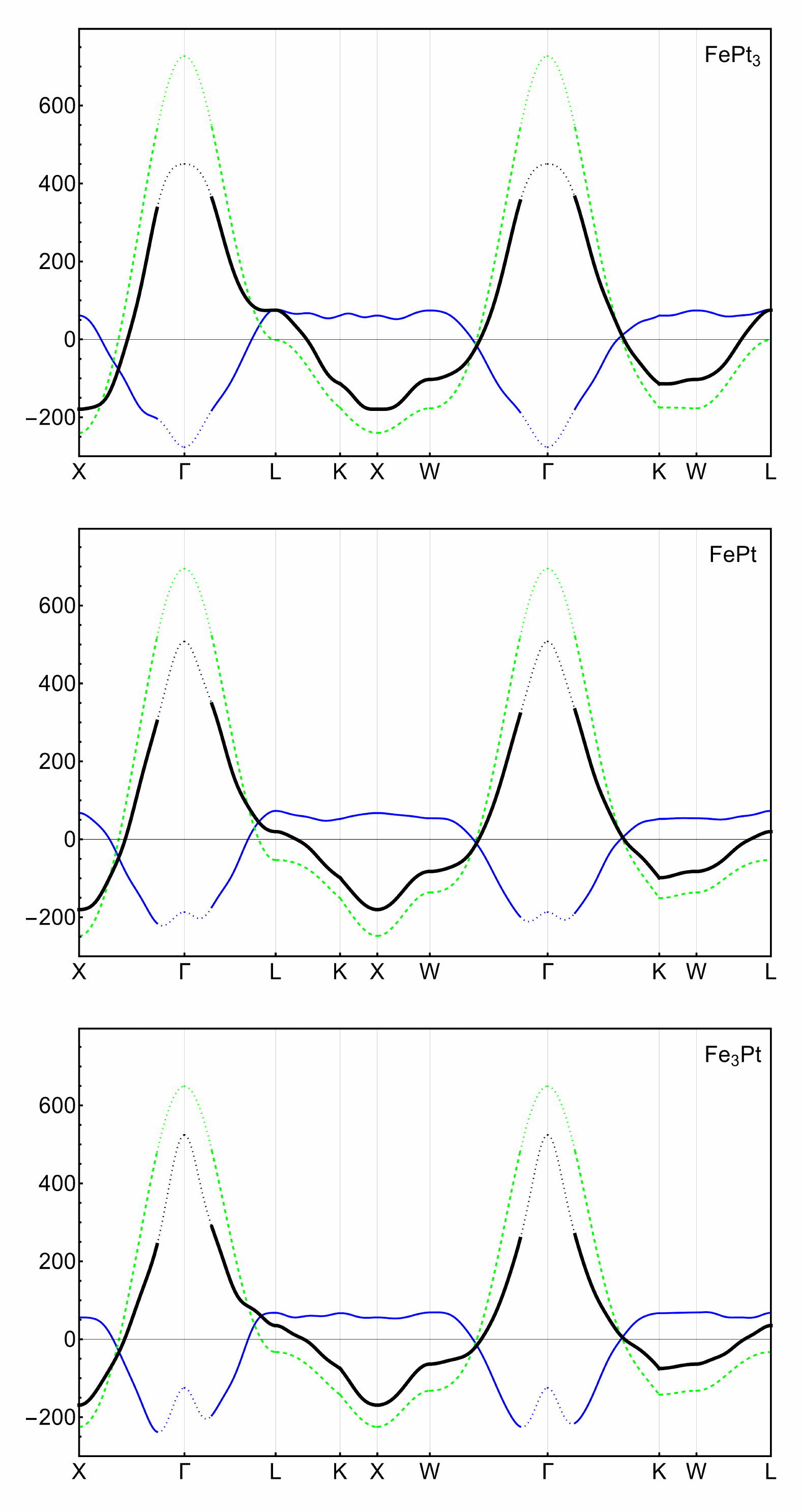}
\caption{(Color online) Effective pair interaction $J_\mathrm{eff}(\mathbf{k})$ (meV units) in Cu-Au and Fe-Pt alloys for small deviations from the random alloy. ``OV'' stands for optimized volume. Dashed green: chemical contribution; solid blue: strain-induced contribution; dotted red: homogeneous strain contribution (only for CuAu); thick black lines: total effective interaction. The dotted segments of the lines near the $\Gamma$ point show the region where the strain-induced contribution is not correctly captured by the auxiliary cluster expansion.}
\label{fig:veff}
\end{figure*}

Figure \ref{fig:veff} shows the Fourier transform $J_\mathrm{eff}(\mathbf{k})$ of the total effective potential along with its strain-induced and chemical parts. $J_\mathrm{eff}(\mathbf{k})$ describes the energy of a concentration wave at a wave vector $\mathbf{k}$ in the random alloy.

Although the data sets used to construct the auxiliary cluster expansions for the relaxation energy include hundreds of structures, the shortest $\mathbf{k}$ vectors represented by these structures are only on the order of 10\% of the reciprocal lattice vector, and the number of structures with such short $\mathbf{k}$ vectors is relatively small. Therefore, the auxiliary cluster expansion fit is insensitive to the behavior of the effective interaction in the vicinity of the $\Gamma$ point. In this small region in the Brillouin zone the interactions are shown by dotted lines in Fig.\ \ref{fig:veff} to emphasize that they are not captured correctly by the fit. On the other hand, ordering tendencies are most sensitive to the behavior of $J_\mathrm{eff}(\mathbf{k})$ at the periphery of the Brillouin zone where it is reliably captured by the fit. In systems under consideration $J_\mathrm{eff}(\mathbf{k})$ reaches its minimum at the zone boundary (at or near the X point), and the inaccuracy of $J_\mathrm{eff}(\mathbf{k})$ near the $\Gamma$ point is unlikely to lead to incorrect predictions for the phase transitions. As it was mentioned above, the situation is different is systems undergoing spinodal decomposition, where it is important to know $J_\mathrm{eff}(\mathbf{k})$ near the $\Gamma$ point.

We can see from Fig.\ \ref{fig:veff} that in most cases the plots of $J_\mathrm{eff}(\mathbf{k})$ for the strain-induced contribution display abrupt turns near the $\Gamma$ point where the fit can no longer capture the $\mathbf{k}$ dependence correctly. These errors are harmless in all systems except CuAu$_3$ where the total effective interaction reaches minimum at the $\Gamma$ point. In this system the auxiliary cluster expansion leads to an unacceptable loss of accuracy.

Inspection of Fig.\ \ref{fig:veff} shows that the overall shape of the total effective interaction is quite similar in Cu$_{0.75}$Au$_{0.25}$ and Cu$_{0.5}$Au$_{0.5}$ systems, as well as in Fe-Pt alloys at all three compositions. The global minimum of $J_\mathrm{eff}(\mathbf{k})$ is reached at the X point in all systems with 50\% and 75\% of the $3d$ element (Cu or Fe). The minimum at the X point in the Cu$_3$Au alloy agrees with the conclusions of Ref.\ \onlinecite{Wolverton}. \cite{note-jeff} The minimum at the X point is sharper in the CuAu system compared to Cu$_3$Au, and in Fe-Pt alloys compared to Cu-Au.
In Fe$_{0.25}$Pt$_{0.75}$ there is a flat region of $J_\mathrm{eff}(\mathbf{k})$ near the X point, with the minimum shifted away from X. This splitting of the minimum at X also does not have immediate consequences, because it is likely to be modified by magnetic disorder (see Section \ref{mcresults} for further discussion).

\section{Monte Carlo simulations}\label{mcresults}

In the previous sections we have been occupied with the construction of configurational Hamiltonians including the relaxation energy, and now we are ready to examine the thermodynamic properties.
The configurational Hamiltonian in our approach corresponds to a fixed overall composition. It is thus potentially suitable for the prediction of coherent phase transformations in a fixed-concentration alloy in a typical quench-and-anneal experiment as long as the thermodynamic simulations are also performed at a fixed composition, i.\ e.\ within the canonical ensemble. As it was mentioned in the introduction, this general scheme (composition-dependent Hamiltonian plus particle-conserving thermodynamic or kinetic simulation) appears to be necessary on general physical grounds for the description of a coherent phase transformation involving phase separation. The procedure developed in this work is, however, not ready for this purpose, because in a phase-separated state the crucial long-range part of the elastic interaction is not captured by the auxiliary cluster expansion. Clearly, a more sophisticated approximation to the CLDM relaxation energy is necessary that would both represent the long-range many-body interaction faithfully and either be computable at the rate required in Monte Carlo simulations or be amenable to a reasonably accurate statistical approximation (such as a suitably adapted cluster variation method). Another issue is the difficulty in describing the homogeneous strain within CLDM, which we have discussed above. As long as the computational cell is filled by all possible domain types of the product phases, the macroscopic symmetry of the disordered phase is not broken. If the latter is cubic, the traceless part of the uniform strain tensor vanishes. In this situation, which is typical at early stages of a kinetic Monte Carlo simulation, the homogeneous strain problem is not of major importance. However, when the domain size becomes comparable with the size of the computational cell (which is typical in Monte Carlo simulations), the homogeneous strain can generally not be ignored. We leave these problems for future studies.

Our present goal is, however, more modest: we are interested in finding the ordering phase transitions at compositions that are close to the points of equal concentration in the phase diagrams. In this situation the long-range part of the interaction is of minor importance, and we expect the auxiliary cluster expansion to be quite adequate. Once the CLDM expression for the relaxation energy has been replaced by the finite-range auxiliary cluster expansion, the choice of the statistical ensemble is of no consequence. Therefore, the Monte Carlo simulations were performed using the efficient semi-grand canonical ensemble as implemented in the ATAT package. \cite{emc2} In this approach the chemical potential difference for the two components is fixed, and the temperature of the phase transition is signalled either by the discontinuity of the concentration or by a singularity in the heat capacity. For the present purposes we select the chemical potential in such a way that the phase transition occurs at the target concentration (i.\ e.\ the same concentration at which the configurational Hamiltonian has been constructed). The discontinuity of the concentration at the target concentrations was found to be small, validating the applicability of the model.

The results of Monte Carlo simulations for Cu-Au and Fe-Pt alloys, are presented in Table \ref{tab:mc}. For each system we list two values of the critical temperature $T_c$: one obtained using only the cluster expansion for $H_{chem}$ and another based on the representation of the full Hamiltonian.

The ordering temperatures $T_c$ for {Fe$_3$Pt} and {FePt} are in good agreement with experimental data. The vibrational entropy neglected here is expected to reduce $T_c$ (in Cu-Au systems this reduction is \cite{Ozolins} about 15\%). Therefore, the agreement with experiment for Fe$_3$Pt is very good, which appears to validate the neglect of the shape relaxation energy. On the other hand, for FePt the model appears to underestimate $T_c$, and for the FePt$_3$ system $T_c$ comes out at half the experimental value. The calculated $T_c$ in FePt is in good agreement with the value of $1514$ K obtained previously \cite{ChB} using Monte Carlo simulations based on the short-range pair interaction which was fitted using the Connolly-Williams structure inversion method.

\begin{table}[hbt]
\caption{Ordering temperatures $T_c$ (K) and phases appearing in Monte Carlo simulations. For FePt$_3$ the boundary of the disordered region at 25\% Cu is listed. For CuAu$_3$ the results are not meaningful (NM).}
\begin{tabular}{|c|c|c|c|c|c|c|}
\hline
 & \multicolumn{2}{c|}{ {Fe$_3$Pt}  } & \multicolumn{2}{c|}{ {FePt}  }   &  \multicolumn{2}{c|}{ {FePt$_3$} } \\
 \cline{2-7}
                            & $T_c$& Phase& $T_c$& Phase&$T_c$& Phase\\
\hline
$H_{chem}$ only             & 1580  &  L1$_2$        & 2000  & L1$_0$         & 1350   &  L1$_2$  \\
\hline
Full    		    		& 1200  & {L1$_2$}       & 1460  & {L1$_0$}       & 840    & {$\beta_2+A1$} \\
\hline
Ref.\ \onlinecite{ChB}      & ---   & ---            & 1514  & L1$_0$         & ---   &  ---      \\
\hline
Experiment \cite{FePtexp}   & 1015  & {L1$_2$}       & 1535  & {L1$_0$}       & 1500  & {L1$_2$}  \\
\hline
\multicolumn{7}{c}{\rule{0pt}{1ex}}\\
\hline
 & \multicolumn{2}{c|}{ {Cu$_3$Au} } & \multicolumn{2}{c|}{ {CuAu}\footnote{Reference lattice at $a=3.819$ \AA\ or 3.897 \AA\ (the latter in brackets).} }   &  \multicolumn{2}{c|}{ {CuAu$_3$} }\\
 \cline{2-7}
                                    & $T_c$ & Phase        & $T_c$ & Phase     & $T_c$ & Phase \\
\hline
$H_{chem}$ only                     & 1630  & L1$_2$       & 1275 (1000) & L1$_0$ & 875 & L1$_2$ \\
\hline
Full                                & 850   & {L1$_2$}     & 670 (740)   & {L1$_0$}  & NM & NM \\
\hline
Ref.\ \onlinecite{deWalle2002}      & 460   & {L1$_2$}     & 430   & {L1$_0$}  & --- & --- \\
\hline
Ref.\ \onlinecite{Ozolins2}         & 530   &{L1$_2$}      & 660   &{L1$_0$}   & 750 & {$\beta_2+A1$}\\
\hline
Experiment \cite{CuAuexp}           & 663   & {L1$_2$}     & 683   & {L1$_0$}  & 473 & {L1$_2$}            \\
\hline
\end{tabular}
\label{tab:mc}
\end{table}

The FePt$_3$ alloy undergoes phase separation, which, strictly speaking, invalidates our procedure involving an auxiliary cluster expansion. However, since the effective interaction $J_\mathbf{eff}(\mathbf{k})$ is very high at the $\Gamma$ point in this system, the ordering tendencies are likely to be correctly reproduced. The emergence of the $\beta_2$ phase agrees with the results of Barabash \emph{et al.} \cite{Barabash} who found that it makes a vertex on the convex hull, while the ferromagnetic L1$_2$ phase lies on a tie-line connecting other phases. (The $\beta_2$ phase was also reported in calculations for Cu-Au, \cite{Ozolins2} Fe-Pd \cite{Barabash}, and Co-Pt \cite{Chepulskii-CoPt} systems.) We found that a small change in the cluster expansion brought about by including input structures in a finite concentration range of 66-86\% stabilizes the L1$_2$ phase and yields an ordering temperature of 740 K. However, the experimental ordering temperature for the L1$_2$ phase is about twice higher (1500 K). This discrepancy (and perhaps also the underestimation of $T_c$ for FePt) can be tentatively attributed to our neglect of magnetic disorder. Indeed, all energies and forces have been calculated in the ferromagnetic state, while the ordering phase transitions occur in the paramagnetic state. The influence of magnetic order on the structural energetics of Fe-Pt alloys was also discussed by Barabash \emph{et al.} \cite{Barabash} who found that antiferromagnetic order places the L1$_2$ FePt$_3$ phase on the convex hull instead of the $\beta_2$ phase. Thus, the stability of the L1$_2$ phase is underestimated. For a more detailed treatment of this problem, the CLDM can be extended by including the dependence of forces and force constants on the magnetic configuration of the alloy. This problem is left for future studies.

Turning to the Cu-Au system and bearing in mind the expected reduction of $T_c$ due to the contribution of vibrational entropy,\cite{Ozolins} we see that the predictions for Cu$_3$Cu and CuAu compositions are quite satisfactory. For CuAu the model based on the optimized volume predicts a 10\% higher transition temperature. This variation may be taken as an indication of the uncertainty built into the model. Indeed, the total energy differences taken from first-principles calculations do not depend on the choice of the reference volume, which only affects their formal partitioning between the chemical and strain-induced parts. It is quite natural that the choice of the optimized reference volume significantly reduces the relative magnitude of the strain-induced interaction, as can be seen from the predicted values of $T_c$ for CuAu in Table \ref{tab:mc} (compare the data labeled ``$H_{chem}$ only'' and ``Full''). The value of 740 K based on the optimized volume should be viewed as more reliable, because this choice reduces the errors due to anharmonicity.

For CuAu$_3$ the simulation predicts phase separation at a temperature above 2000 K, which is not meaningful and clearly due to the failure of the auxiliary cluster expansion to capture the long-range part of the interaction. On the other hand, as mentioned in the Introduction, the semi-grand canonical Monte Carlo simulation is also problematic for a system undergoing coherent phase separation. Since the GGA, in fact, fails to predict the ground state in this system, we did not attempt to improve its description. We have, however, checked that the CLDM, in good agreement with first-principles calculations, predicts that several structures in the input set have energies below that of L1$_2$. It has been recently shown that the use of hybrid exchange-correlation functionals eliminates this failure of semi-local functionals, \cite{Zhang} which may lead to a better description of the Au-rich side of the phase diagram.

Comparison of the transition temperatures obtained including (lines labeled ``Full'' in Table \ref{tab:mc}) and neglecting (lines labeled ``$H_{chem}$ only'') the strain-induced interaction illustrates the relative magnitude of the latter. For example, in FePt it reduces the predicted $T_c$ by 27\%. This reduction agrees very well with a rough estimate that could be made based on the results of first-principles calculations. Indeed, the relaxation energy of the Fe$_{0.5}$Pt$_{0.5}$ SQS16 structure is 38 meV, while the energy difference between the unrelaxed SQS16 and L1$_0$ structures is 135 meV; thus the $T_c$ reduction could be roughly predicted to be 28\%. Note that the average volume relaxation in FePt is only 1.3 meV (see Table \ref{tab:VShRel}), and therefore the relative magnitude of the strain-induced interaction can not be materially reduced by selecting a different reference volume. On the other hand, in Ref.\ \onlinecite{ChB} the traditional KKKM was employed to estimate the strain-induced interaction in FePt using the elastic constants of Pt and the concentration expansion coefficient. The strain-induced interaction was found to be very small, only reducing $T_c$ by about 60 K. This drastic failure of KKKM shows that its uncontrolled approximations make it unsuitable even for rough estimates of the strain-induced interaction in concentrated $3d$-$5d$ alloys.

\section{Conclusions}\label{conclusions}

We have presented a configuration-dependent lattice deformation model (CLDM) designed to represent the strain-induced interaction in concentrated size-mismatched alloys, which systematically generalizes the Kanzaki-Krivoglaz-Khachaturyan formalism. Our present treatment is done within the harmonic approximation. Both the Kanzaki forces and force constants referenced from the undistorted lattice are given by many-body cluster expansions constructed based on first-principles calculations. This model has been applied to Cu-Au and Fe-Pt alloys, treating three compositions near 25\%, 50\%, and 75\% as separate systems. Large asymmetry between the two components leads to a strongly non-pairwise strain-induced interaction in all systems. The ability to capture these singularly long-range many-body interactions is the main advantage of CLDM.

The model was found to provide a rather accurate representation of the relaxation energy under the restriction of constant uniform strain, with the main source of error being the neglect of anharmonicity in strongly relaxing configurations. However, the configuration-dependent stress is not adequately represented by the models, which may be due to the slow real-space convergence of the corresponding lattice sums. As a result, the contribution of striction (i.\ e.\ relaxation of the unit cell shape) must be included separately when it is important.

The phonon spectra of random alloys based on the configurationally-dependent force constants are in good qualitative agreement with the available experimental data, but an accurate calculation of the vibrational entropy is not possible without the inclusion of anharmonic terms.

For further applications, the CLDM relaxation energy is refitted to a multi-parameter real-space cluster expansion using several thousand input structures. The effective pair interaction $J_\mathbf{eff}(\mathbf{k})$ obtained from this fit (Fig.\ \ref{fig:veff}) can be used to analyze the ordering tendencies in a disordered alloy.

The calculated ordering transitions (Table \ref{tab:mc}) are in good agreement with experiment with the exception of FePt$_3$ where the stability of the L1$_2$ phase is underestimated due to the neglect of magnetic disorder, and CuAu$_3$ due to the known failure of GGA to predict the correct ground state.

Overall, the CLDM provides an accurate and practical approach for the description of strain-induced interaction in concentrated alloys. Some natural future developments may include the incorporation of spin-dependent forces and force constants in magnetic systems to describe the interplay between structural and magnetic disorder, and the inclusion of anharmonic terms to improve the prediction of relaxation energies and to enable the calculation of the vibrational entropy. The reasons for the failure of the current models to represent the configuration-dependent stress tensor also require further analysis.

\acknowledgments

We thank Andrei Ruban, Sergey Barabash, and Igor Abrikosov for useful discussions. This work was supported by the DOE EPSCoR State and National Laboratory Partnership Program (Grant No.\ DE-SC0001269) and by National Science Foundation (Grant No.\ DMR-1308751). Computations were performed utilizing the Holland Computing Center of the University of Nebraska.


\begin{thebibliography}{99}
\bibitem{deFontaine} D.\ de Fontaine, Solid State Physics \textbf{34}, 73 (1979); \emph{ibid.} \textbf{47}, 33 (1994).

\bibitem{Sanchez} J. M. Sanchez, F. Ducastelle, and D. Gratias, Physica A \textbf{128}, 334 (1984).

\bibitem{Duc} F. Ducastelle, Order and Phase Stability in Alloys, Elsevier Science, New York, 1991.

\bibitem{Zunger} A. Zunger, in NATO ASI on Statics and Dynamics of Alloy Phase Transformation,
Vol.\ 319, ed.\ P.\ E.\ Turchi and A.\ Gonis, Plenum Press, New York, 1994, p. 361.

\bibitem{Walle} A. van de Walle and G. Ceder, J. Phase Equil. \textbf{23}, 348 (2002).

\bibitem{Ruban-review} A.\ V.\ Ruban and I.\ A.\ Abrikosov, Rep.\ Prog.\ Phys.\ \textbf{71}, 046501 (2008).

\bibitem{Krivoglaz} M.\ A.\ Krivoglaz, X-ray and neutron diffraction in nonideal crystals (Springer, Berlin, 1996).

\bibitem{Khachaturyan} A.\ G.\ Khachaturyan, Theory of structural transformations in solids (Wiley, New York, 1983).

\bibitem{CW} J. W. D. Connolly and A. R. Williams, Phys. Rev. B \textbf{27}, 5169 (1983).

\bibitem{Laks} D.\ B.\ Laks, L.\ G.\ Ferreira, S.\ Froyen, and A.\ Zunger, Phys.\ Rev.\ B \textbf{46}, 12587 (1992).

\bibitem{Ozolins2} V.\ Ozoli\c{n}\v{s}, C.\ Wolverton, and A.\ Zunger, Phys.\ Rev.\ B \textbf{57}, 6427 (1998).

\bibitem{Blum} V.\ Blum and A.\ Zunger, Phys.\ Rev.\ B \textbf{70}, 155108 (2004).

\bibitem{Matsubara} T.\ J.\ Matsubara, J.\ Phys.\ Soc.\ Jpn \textbf{7}, 270 (1952).

\bibitem{Kanzaki} H.\ Kanzaki, J.\ Phys.\ Chem.\ Solids \textbf{2}, 24 (1957).

\bibitem{Ruban-2010} A.\ V.\ Ruban, V.\ I.\ Baykov, B.\ Johansson, V.\ V.\ Dmitriev, and M.\ S.\ Blanter, Phys.\ Rev.\ B \textbf{82}, 134110 (2010).

\bibitem{Ruban-2014} A.\ V.\ Ruban, V.\ A.\ Popov, V.\ K.\ Portnoi, and V.\ I.\ Bogdanoff, Phil.\ Mag.\ \textbf{94}, 20 (2014).

\bibitem{vdW-RMP} A.\ van de Walle and G.\ Ceder, Rev.\ Mod.\ Phys.\ \textbf{74}, 11 (2002).

\bibitem{Fultz} B.\ Fultz, Prog.\ Mater.\ Sci.\ \textbf{55}, 247 (2010).

\bibitem{Shirley} A.\ I.\ Shirley and C.\ K.\ Hall, Phys.\ Rev.\ B \textbf{33}, 8084 (1986); \textbf{33}, 8099 (1986).

\bibitem{BPSV} K.\ D.\ Belashchenko, I.\ R.\ Pankratov, G.\ D.\ Samolyuk, and V.\ G.\ Vaks, J.\ Phys.: Condens.\ Matter \textbf{14}, 565 (2002).

\bibitem{Shchyglo} O.\ Shchyglo, A.\ D\'iaz-Ortiz, A.\ Udyansky, V.\ N.\ Bugaev, H.\ Reichert, H.\ Dosch, and R.\ Drautz, J.\ Phys.:\ Condens.\ Matter \textbf{20}, 045207 (2008).

\bibitem{Ruban-tetra} A.\ V.\ Ruban, S.\ I.\ Simak, S.\ Shallcross, and H.\ L.\ Skriver, Phys.\ Rev.\ B \textbf{67}, 214302 (2003).

\bibitem{Williams1} R.\ O.\ Williams, Metall.\ Trans.\ A \textbf{11}, 247 (1980).

\bibitem{Williams2} R.\ O.\ Williams, Calphad \textbf{8}, 1 (1984).

\bibitem{CL} J.\ W.\ Cahn and F.\ Larch\'e, Acta Metall.\ \textbf{32}, 1915 (1984).

\bibitem{Larche} F.\ C.\ Larch\'e, Annu.\ Rev.\ Mater.\ Sci.\ \textbf{20}, 83 (1990).

\bibitem{Johnson} W.\ C.\ Johnson, Metall.\ Trans.\ A \textbf{18}, 1093 (1991).

\bibitem{Bloechl} P.\ E.\ Bl\"ochl, Phys.\ Rev.\ B \textbf{50}, 17953 (1994).

\bibitem{VASP-PAW} G.\ Kresse and D.\ Joubert, Phys.\ Rev.\ B \textbf{59}, 1758 (1999).

\bibitem{PBEsol} J.\ P.\ Perdew, A.\ Ruzsinszky, G.\ I.\ Csonka, O.\ A.\ Vydrov, G.\ E.\ Scuseria, L.\ A.\ Constantin, X.\ Zhou, and K.\ Burke, Phys.\ Rev.\ Lett.\ \textbf{100}, 136406 (2008).

\bibitem{VASP} G.\ Kresse and J.\ Hafner, Phys.\ Rev.\ B \textbf{48}, 13115 (1993); G.\ Kresse and J.\ Furthm\"uller, Comput.\ Mater.\ Sci. \textbf{6}, 15 (1996); Phys.\ Rev.\ B \textbf{54}, 11169 (1996).

\bibitem{Born} M.\ Born and and K.\ Huang, Dynamical Theory of Crystal
Lattices (Oxford University Press, London, 1956).

\bibitem{Triguero} C.\ Triguero, M.\ Porta, and A.\ Planes, Phys.\ Rev.\ B \textbf{73}, 054401 (2006).

\bibitem{Chepulskii} R. V. Chepulskii, S. V. Barabash, and A. Zunger, Phys.\ Rev.\ B \textbf{85}, 144201 (2012).

\bibitem{TDEP} O.\ Hellman and I.\ A.\ Abrikosov, Phys.\ Rev.\ B \textbf{88}, 144301 (2013).

\bibitem{Bogdanoff-03} P.\ D.\ Bogdanoff, T.\ L.\ Swan-Wood, and B.\ Fultz, Phys.\ Rev.\ B \textbf{68}, 014301 (2003).

\bibitem{Yue} A.\ F.\ Yue \emph{et al.}, Hyperfine Interactions \textbf{141/142}, 249 (2002).

\bibitem{Noda} Y.\ Noda and Y.\ Endoh, J.\ Phys.\ Soc.\ Jpn \textbf{57}, 4225 (1988).

\bibitem{Wolverton} C.\ Wolverton, V.\ Ozoli\c{n}\v{s}, and A.\ Zunger, Phys.\ Rev.\ B \textbf{57}, 4332 (1998).

\bibitem{note-jeff} Note that $J_\mathrm{eff}(\mathbf{k})$ for Cu$_3$Au calculated from Eq.\ (\ref{jeff}) can not be directly compared to $J(\mathbf{k})$ shown in Fig.\ 7 of Ref.\ \onlinecite{Wolverton}, where the many-body contribution to the energy was linearly interpolated between the L1$_2$ and D0$_{22}$ phases. This interpolation fails to give a zero derivative at the W point along the $\langle 1\zeta0\rangle$ line as required by symmetry.

\bibitem{emc2} A.\ van de Walle and M.\ Asta, Modelling Simul.\ Mater.\ Sci.\ Eng.\ \textbf{10}, 521 (2002).

\bibitem{Ozolins} V.\ Ozoli\c{n}\v{s}, C.\ Wolverton, and A.\ Zunger, Phys.\ Rev.\ B \textbf{58}, R5897(R) (1998).

\bibitem{ChB} R.\ V.\ Chepulskii and W.\ H.\ Butler, Phys.\ Rev.\ B \textbf{72}, 134205 (2005).

\bibitem{Barabash}  S.\ V.\ Barabash, R.\ V.\ Chepulskii, V.\ Blum, and A.\ Zunger, Phys.\ Rev.\ B \textbf{80}, 220201 (2009).

\bibitem{Chepulskii-CoPt} R.\ V.\ Chepulskii and S.\ Curtarolo, Appl.\ Phys.\ Lett.\ \textbf{99}, 261902 (2011).

\bibitem{FePtexp} H. Okamoto, in Phase Diagrams of Binary Iron Alloys, ed.\ H.\ Okamoto, ASM International (Materials Park, OH, 1993), pp. 330-36.

\bibitem{deWalle2002} A.\ van\ de\ Walle, G.\ Ceder,  Journal\ of\ Phase\ Equilibria  Vol. \textbf{23}, 348 (2002)

\bibitem{CuAuexp}  H. Okamoto, D. J. Chakrabarti, D. E. Laughlin, and T. B. Massalski, Bull. Alloy Phase Diagrams \textbf{8}, 454 (1987).

\bibitem{Zhang}    Y.\ Zhang, G.\ Kresse, and C.\ Wolverton, Phys.\ Rev.\ Lett. \textbf{112}, 075502 (2014).

\end{thebibliography}
\end{document}